\documentclass{jfm}
\usepackage{graphicx}
\usepackage[]{amsmath}
\usepackage[]{xcolor}
\usepackage[]{subfigure}
\usepackage{upgreek}
\usepackage{overpic}
\usepackage[]{multirow}
\usepackage[]{url}
\usepackage[]{nicefrac}
\usepackage{bm}
\usepackage[]{graphicx}
\usepackage{booktabs}  %  引入三线表宏包
\usepackage{natbib}
\usepackage{subfigure}
\usepackage{makecell}

\shorttitle{Artificial intelligence control of a turbulent jet}

\title{Artificial intelligence control\\ of a turbulent jet}

\author{Yu Zhou\aff{1}\corresp{\email{yuzhou@hit.edu.cn}},
Dewei Fan\aff{1},
Bingfu Zhang\aff{1},
Ruiying Li\aff{2}\footnote{Present address: Renault Group 1, Avenue du Golf 78280 Guyancourt}\\ and
Bernd R.~Noack\aff{1,3}\corresp{\email{bernd.noack@hit.edu.cn}}}

\affiliation{
\aff{1}
Institute for Turbulence-Noise-Vibration Interaction and Control,
Harbin Institute of Technology (Shenzhen),
People's Republic of China
\aff{2}
Institut PPRIME,
CNRS -- Universit\'e de Poitiers -- ISAE-ENSMA,
86962 Futuroscope Chasseneuil, France
\aff{3}
Institut f\"ur Str\"omungsmechanik und Technische Akustik (ISTA),
Technische Universit\"at Berlin,
M\"uller-Breslau-Stra{\ss}e 8,
D-10623 Berlin, Germany
}

\begin{document}

\maketitle
\begin{abstract}
An artificial intelligence (AI) control system 
is developed to maximize the mixing rate of a turbulent jet. 
This system comprises six independently operated unsteady minijet actuators, 
two hot-wire sensors placed in the jet, 
and genetic programming for the unsupervised learning of a near-optimal control law. 
The ansatz of this law includes multi-frequency open-loop forcing, 
sensor-feedback and nonlinear combinations thereof. 
Mixing performance is quantified by the decay rate of the centreline mean velocity of jet. 
Intriguingly, the learning process of AI control discovers the classical forcings, 
i.e. axisymmetric, helical and flapping achievable from conventional control techniques, 
one by one in the order of increased performance, 
and finally converges to a hitherto unexplored forcing. 
Careful examination of the control landscape unveils typical control laws, 
generated in the learning process, and their evolutions. 
The best AI forcing produces a complex turbulent flow structure 
that is characterized by periodically generated mushroom structures, 
helical motion and oscillating jet column, 
all enhancing the mixing rate and vastly outperforming others. 
Being never reported before, this flow structure is examined in various aspects, 
including the velocity spectra, mean and fluctuating velocity fields 
and their downstream evolution, and flow visualization images in three orthogonal planes, 
all compared with other classical flow structures. 
Along with the knowledge of the minijet-produced flow and its effect on the initial condition of the main jet, 
these aspects cast valuable insight into the physics behind the highly effective mixing of this newly found flow structure. 
The results point to the great potential of AI in conquering 
the vast opportunity space of control laws 
for many actuators and sensors and in optimizing turbulence.

\end{abstract}

\begin{keywords}
 Artificial intelligence, turbulence, jets, flow control.
\end{keywords}

\maketitle

%\rule{0pt}{0pt}\newpage
%\tableofcontents \newpage
%\setcounter{page}{1}

\def\degree{${}^{\circ}$}

\section{Introduction}
\label{ToC:Introduction}

The turbulent jet is one of classical shear flows discussed in virtually every textbook. 
Its control finds important industrial applications, 
including dilution jets in combustors, 
fuel injection of combustion engines, 
noise mitigation of sub- and supersonic jets 
for civil and military aircrafts, thrust augmenting ejectors, 
thrust vector control, etc. 
The key to control entrainment and mixing processes in a turbulent jet is, 
as in other shear layers, to manipulate the coherent motions. 
When a jet issues from a round nozzle, 
a free shear layer is formed from the nozzle lip and develops downstream. 
Kelvin-Helmholtz instability \citep{Ho1984arfm} inherent in the shear layer rapidly grows, 
resulting in the formation of axisymmetric ring vortices. 
The vortices, along with their subsequent interaction 
(e.g. merging and breakdown), dominate the shear-layer growth and entrainment \citep{crow1971jfm}. 
Shortly downstream of the nozzle exit, three-dimensionality becomes an important feature of the flow structure; 
streamwise vorticity contributes predominantly the entrainment of fluid from the surroundings \citep{liepmann1992jfm}. 
These motions, formed near the nozzle exit, are featured by a wide range of scales, 
varying convection velocity and a rich set of three-dimensional patterns \citep{Garnaud2013jfm}; 
they are sensitive to initial conditions 
(e.g. the turbulence level, boundary layer thickness, nozzle geometry) 
and external periodic disturbances \citep{Vlasov1973fd}, thus highly susceptible for control. 

Jet control can be active or passive. Passive control involves 
a change in geometry such as tabs (e.g. \citet{bradbury1975jfm}), 
non-circular nozzles (e.g. \citet{Husain1983pof}) and chevron nozzles (e.g. \citet{alkislar2007jfm}). 
Although often highly effective, passive techniques are characterized by permanent fixtures. 
Once mounted, tabs are difficult to be relocated. 
Likewise, it is impractical for any engineering application 
to implement frequently noncircular nozzle geometry alteration 
due to cost and physical constraints. 
Furthermore, there are other penalties, e.g. thrust loss and drag. 
Active control requires the input of external power, 
e.g. acoustic excitation (e.g. \citet{Zaman1981jfm}), 
piezo-electric actuators (e.g. \citet{wiltse1993jfm}), 
plasma actuators (e.g. \citet{Samimy2007jfm}), 
synthetic jet (e.g. \citet{Tamburello2006jot}), 
flip-flop jets (e.g. \citet{Raman1993AIAA}) 
and steady/unsteady minijets \citep{Zhou2012AIAA, Yang2016jfm}. 
The active method has potential to achieve more flexible and drastic flow modifications, 
which is a great advantage over the passive 
(e.g. \citet{zaman1994pof}, \citet{Longmire1996pof}, 
\citet{reeder1996jfm}).

Many active control studies of turbulent jets involve 
the open-loop periodic forcing of a pre-specified form, 
e.g. axisymmetric, flapping or helical forcing. 
\citet{Broze1994jfm} deployed four speakers upstream of the nozzle 
to add a longitudinal component of perturbation to the mean flow. 
The acoustic source produced axisymmetric forcing which was found 
to amplify vortex ring structures and subsequent vortex pairing. 
\citet{Koch1989pof} generated helical forcing 
on a turbulent round air jet using four speakers, 
each being $90^{\circ}$ out of phase with the adjacent speaker.
\citet{Yang2016Turbulent} used two unsteady radial minijets separated by $60^{\circ}$ or $120^{\circ}$ to produce flapping jet column, 
which enhanced greatly jet mixing. 
The combination of individual forcings is also investigated. 
\citet{Juvet1987} optimized experimentally the combinations 
of axisymmetric axial and helical forcing to augment mixing. 
The axial excitation was produced by a loudspeaker placed upstream 
of the jet and perpendicularly to the centreline. 
The helical excitation was generated by four external loudspeakers. 
Their acoustic excitations were directed via waveguides 
at an angle around the jet exit lip to the shear layer 
where the flow is most sensitive to acoustic forcing. 
While the axial excitation led to the early formation of large-scale vortices, 
the helical caused the shear layer to roll up into staggered vortex structures. 
This combined excitation generated a bifurcating jet 
with a much larger spreading angle than the single excitation 
when the frequency ratio of the axial to that 
of the helical excitation was equal to 2 \citep{Reynolds2003arfm}. 
Hilgers \& Boersma's (2001) three-dimensional direct numerical simulation of a turbulent jet demonstrated 
that the superposition of two counter-rotating helical modes of the same excitation frequency $f_e$ and 
one axial excitation of 2$f_e$ produced a bifurcating jet 
whose centreline mean velocity and scalar concentration decayed 
faster than those of the counter-rotating helical actuation alone. 

\citet{Tyliszczak2015JOT} and \citet{Tyliszczak2018ijhff} simulated highly mixed multi-armed bifurcating jets using axial and helical excitations. 
However, it would be very difficult or time-consuming 
for conventional active controls to find the globally optimal solution 
for the combined excitations where many control parameters 
are generally involved. 
For instance, the control optimization of a turbulent jet has so far 
typically involved up to two control parameters, 
such as the actuation amplitude and frequency. 
Then, the optimization of combined modes, like axisymmetric forcing and flap-ping forcing, 
may involve at least four independent control parameters, 
i.e. the amplitude and frequency of each mode 
(e.g. \citet{Hilgers2001fdr}). 
The search for its optimal solution is then already a challenge. 
If the control parameters for each mode is increased to three or four 
such as the amplitude, frequency, duty cycle and diameter ratio 
of an unsteady jet (e.g. \citet{ARUN2018jfm}) or 
multiple independent actuators are deployed, the search for 
the globally optimal solution of the combined modes would be a daunting task. 
\citet{Koumoutsakos2001AIAA} and \citet{Hilgers2001fdr} have pioneered the jet mixing optimization 
with 3 and 4 actuation parameters using Rechenberg’s \citeyear{Rechenberg1973book} evolutionary strategy.

Model-based control comes, if doable, with the deep understanding of actuation dynamics, 
regardless of open or closed loops. 
In simulations, the linear dynamics can be accurately resolved 
by discretized Navier-Stokes (N-S) equations 
\citep{Kim2007arfm, Sipp2010amr}. 
In experiments, linear stochastic estimation \citep{Tinney2006OEIF} 
has been successfully applied to resolve the flow physics 
from measurement signals and PIV measurements. 
The linearized N-S dynamics can be encapsulated in reduced-order models employing several dominant non-normal global stability eigenmodes. 
The downstream evolution of wave-packets can be real-time estimated 
in a high-Reynolds-number turbulent jet 
thanks to the development of transfer functions 
based on the parabolized stability equations \citep{sasaki2017jfm}. 
So can the closed-loop control of fluctuations in a low-Reynolds-number shear layer \citep{Sasaki2018tcfd}. 
These control-oriented models have significantly contributed to the understanding of the manipulated jet dynamics. 

Model-free approaches may yield performance benefits 
from nonlinear dynamics which is too complex for control-oriented models. 
A new model-free self-learning approach for 
general non-linear control laws has been developed 
by \citet{Dracopoulos1997book} 
for commanding satellite motion and was 
re-discovered in fluid mechanics as 
Machine Learning Control or MLC \citep{gautier2015jfm}. 
A review of dozens of MLC experiments and simulations is provided 
by \citet{Noack2019fssic}. 
The first MLC experiment was set to enhancing shear layer mixing 
with 96 jet actuators driven in unison and 25 hot-wire sensors 
for feedback control \citep{Parezanovic2016jfm}. 
The optimization of shear-layer mixing resulted in destabilizing phasor control, 
i.e. the feedback excitation of the dominant frequency. 
The control enhanced and synchronized downstream large-scale vortices 
with a frequency of one sixth of the initial Kelvin-Helmholtz (K-H) instability. 
\citet{Li2017aom} deployed four Coanda jet actuators placed 
at the trailing edge of an Ahmed body and 12 pressure sensors 
at the back side, achieving a drag reduction by 22\% 
(where about 10\% was attributed to the Coanda effect) 
when the excitation frequency was much higher than the predominant and even the shear-layer frequencies in the wake. 
Using an unsteady single jet actuator driven by an online PIV-based sensing, 
\citet{gautier2015jfm} cut short the reattachment length of flow over a backward-facing step. 
They observed surprisingly the enhancement of a low-frequency flapping mode, 
instead of the excitement of the dominant K-H vortex shedding. 
MLC has matched with or outperformed existing control strategies 
and solved the combined task of picking the nonlinear mechanisms 
for performance optimization and selecting the best sensors. 
These model-free control studies show that the actuation mechanism 
can be very difficult to anticipate, 
thus implying a challenge to any model-based control. 

It is worth pointing out that MLC has never been applied 
to multiple independently operated actuators, 
resembling a distributed actuation, in experiments so far. 
MLC laws, previously developed, 
have been of small to moderate complexity, 
e.g. the phasor control, threshold-level based control, 
periodic or two-frequency forcing \citep{Duriez2016mlc}, 
as the actuators are typically driven by a single actuation command. 
Indeed, the use of independent actuators may increase dramatically the level of control complexity. 
For example, assume that one unsteady minijet, used to maximize jet mixing, involves three parameters, 
i.e. the actuation frequency $f_a$, velocity $U_a$ and duty cycle $\alpha$. 
Then, if the number is increased to up to say six independent minijets 
spatially distributed around main jet, the independent control parameters will be tremendously increased. 
Then one naturally wonders what the globally optimal solution 
of the problem is and whether an AI control technique 
could be developed to find this solution. 
Furthermore, what turbulent flow structure might this global optimal solution or forcing produce? 

This experimental work aims to address the issues raised above 
and to optimize jet entrainment/mixing with six independently unsteady minijets 
placed upstream of nozzle exit, extending the MLC jet control 
using a single unsteady minijet \citep{Wu2018eif}. 
The manuscript is organized as follows. 
The experimental setup and minijet-produced flow, 
along with its effect on the jet initial conditions, 
are described in \S~\ref{ToC:Exp_details} and \S~\ref{ToC:Minijet actuation}, respectively. 
The following three sections \ref{ToC:AIControl_system}, 
\ref{ToC:Learningcontrollaw} and \ref{ToC:Flow aspect} describe 
the AI control system developed, the outcome of the AI-based learning 
and the resulting turbulent flow structures. 
The work is concluded in \S~\ref{ToC:Conclusions}.

%***********************************************************************
\section{Experimental details}
\label{ToC:Exp_details}
\subsection{Jet facility}
\label{ToC:Exp_setup}
 %--- Figure ------------------------------------------------------------
\begin{figure}
\begin{center}
\includegraphics[width=0.85\textwidth]{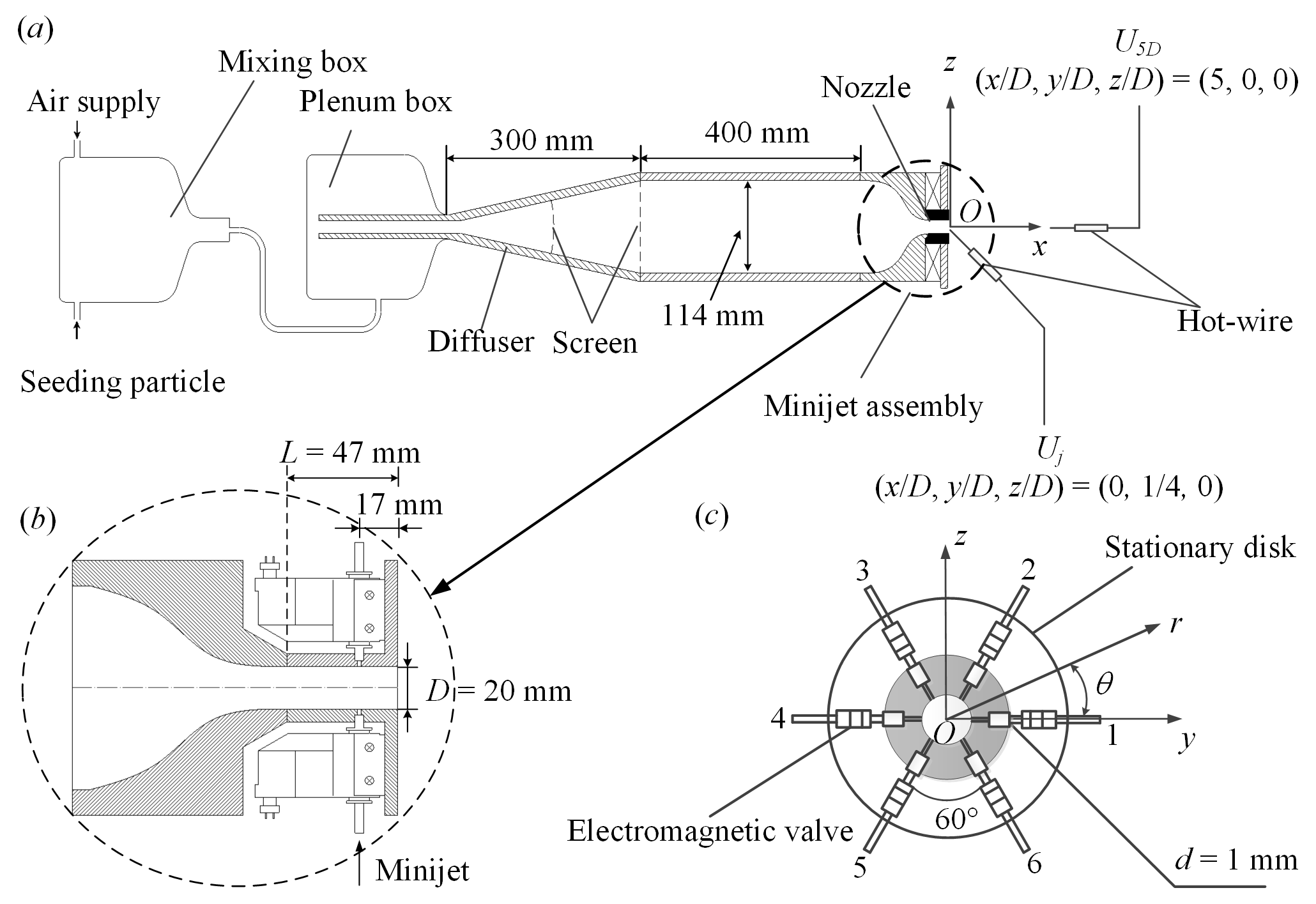} \\
\end{center}
\caption{Sketch of the experimental setup:
($a$) main jet facility;
($b$) minijet assembly;
($c$) minijet arrangement.}
\label{Fig:Setup}
\end{figure}
 %---------------------------------------------------------------

Experiments were conducted in a round air jet facility, 
as schematically shown in figure \ref{Fig:Setup}. 
The facility was placed in an air-conditioned laboratory 
where the room temperature remains constant within $\pm 0.5^{\circ}$C, 
centrally deployed in an area of approximately 2.5 m in width and 2 m in height, enclosed by fabric walls. 
In order to minimize the effects of the wall on the jet, 
the nozzle exit is 4.0 m away from the fabric partition wall 
and the distance is well over 70 times jet exit diameter required 
for neglecting the wall effects \citep{Malmstr1997jfm}. 
As the jet is highly sensitive to background noise, 
careful measures are taken to avoid any external interference to airflow. 

The compressed air of the round jet comes from a constant 5 bar gauge pressure, 
mixed with seeding particles in the mixing chamber 
in the case of the particle image velocimetry (PIV) 
or flow visualization measurements, 
and then enters into a plenum chamber, 
composed of a 300 mm long diffuser of 15$^{\circ}$ in half-angle 
and a 400 mm long cylindrical settling chamber with an inner diameter of 114 mm. 
The flow passes two screens before entering the smooth contraction nozzle \citep{ARUN2018jfm}, 
which is extended by a 47 mm long smooth tube of the same diameter 
as the nozzle exit $D$ (= 20 mm). 
The Reynolds number $Re_D$ = $\overline{U}_jD/\nu$ of the main jet 
is fixed at 8000, where $U_j$ is the centreline velocity measured 
at the nozzle exit, the overbar denotes time-averaging and $\nu$ is the kinematic viscosity of air. 
A Cartesian coordinate system ($x, y, z$) is defined in figure \ref{Fig:Setup}($a$, $c$), 
with its origin at the centre of the jet exit and the $x$-axis pointing in the direction of flow. 
Measurements were conducted in the $x-z$, $x-y$ and $y-z$ planes of the main jet. 
The instantaneous and fluctuating velocities in the $x$, $y$ and $z$ directions 
are denoted by ($U, V, W$) and ($u, v, w$), respectively.

Six unsteady control minijets issued from orifices with a diameter 
of 1 mm are equidistantly placed around the extension tube 
at $x_i = -0.85 D$, $y_i = (D/2) \cos \theta_i$, $zi = (D/2) \theta_i$, 
where  $\theta_i$ = $(i-1)2\pi/6$, $i = 1, 2\ldots, 6 $ 
(figure \ref{Fig:Setup}$b, c$). 
Their mass flow rate is determined by a flow-limiting valve 
and monitored by a mass flow meter, 
with a measurement uncertainty of 1 \%, 
and the frequencies and duty cycles are independently controlled 
by individual electromagnetic valves that are operated in an ON/OFF mode. 
The maximum operating frequency of the valves is 500 Hz, 
exceeding three times the preferred-mode frequency, $f_0$ = 135 Hz, 
of the unforced jet, the corresponding Strouhal number 
being $S_t$ = $f_0D/\overline U_j$ = 0.45, 
where $f_0$ is obtained from the power spectral density function 
$E_u$ of streamwise fluctuating velocity $u$ measured 
in the absence of control \cite{Yang2016jfm}.

\subsection{Flow measurements}
\label{ToC:Measurements}

The fluctuating flow velocities are monitored by two tungsten wire sensors of 5 $\mu$m in diameter, 
operated on a constant temperature circuit (Dantec Streamline) 
at an overheat ratio of 0.6, 
one placed at ($x/D, y/D, z/D$) = (0, 1/4, 0) 
and the other at ($x/D, y/D, z/D$) = (5, 0, 0). 
The time-averaged velocity at the latter position is denoted by $U_{5D}$. 
This choice is based on following considerations. 
Firstly, \citet{Zhou2012AIAA} demonstrated that thus defined $K$ is correlated approximately linearly with an equivalent jet half width $R_{eq}$ = $[R_H R_V]^{0.5}$, 
where $R_H$ and $R_V$ are the jet half-widths in two orthogonal planes, 
implying that $K$ is directly connected to the entrainment rate of the manipulated jet. 
Secondly, \citet{Fan2017AIAA} found that the difference $\triangle K$ 
between the $K$ values with and without control reaches the maximum 
at $x/D \approx 5$, 
that is, the centreline mean streamwise velocity measured 
at $x/D = 5$ is most sensitive to the change in the control parameters. 
Finally, the variation in $K$ is almost linear from $x/D$ = 0 to $x/D \approx 7$ under control (Fig 7, \citet{Fan2017AIAA}), 
that is, a single value of $K$ may be used 
to describe reasonably well the jet decay rate in the near field under control. 
Both hot-wires are calibrated at the jet exit 
using a pitot-static tube connected to a micromanometer (Furness Controls FCO510). 
The cutoff and sampling frequencies are 3 kHz and 6 kHz for open-loop control experiments, respectively. 
The experimental uncertainty of the hot-wire measurement is estimated to be less than 2\%.

A planar high-speed PIV system, 
with a high-speed camera (Dantec Speed Sence 90C10, 2056 $\times$ 2056 pixels resolution) 
and a pulsed laser source (Litron LDY304-PIV, Nd: YLF, 30 mJ/pulse), 
is deployed for velocity field measurements in the $x-z$, $x-y$ and $y-z$ planes. 
An oil droplet generator (TSI MCM-30) is used to generate fog 
from olive oil with an averaged particle size of 1 $\mu$m for flow seeding. 
Flow illumination is provided by a laser sheet of 1 mm 
in thickness generated by the pulsed laser via a cylindrical lens. 
For velocity measurements in the $x-z$ and $x-y$ planes, 
the captured image covers the area of $x/D \in [0, 6]$ and $y/D$, $z/D \in [-2, 2]$. 
The longitudinal and lateral image magnifications are identical, 0.09 mm per pixel. 
The time interval between two consecutive images is presently chosen to be 25 $\mu$s, 
which is found to yield satisfactory results. 
There are 253$\times$253 velocity vectors, 
the same for the two planes. 
A total of 200 pairs of flow images are captured at a sampling rate of 405 Hz for each set of PIV data. 
In post-processing, a built-in adaptive correlation function of the Flow Map Processor (PIV 2001 type) 
is applied with an interrogation window of 32 $\times$ 32 pixels and a 75\% overlap along both directions. 

The same PIV system is used for flow visualization 
in the three orthogonal planes. 
So are the seeding particles, though their concentration 
is higher than in the PIV measurements 
to provide a clear picture for the flow structure. 
The captured images cover the area of $x/D \in [0, 6]$ and $y/D$ or $z/D \in [-2, 2]$ in the $x-y$ and $x-z$ planes and the area of $y/D$ = $z/D \in [-2, 2]$ at $x/D$ = 0.25 in the $y-z$ plane.

\subsection{Real-time system}
\label{ToC:Real-time}

A National Instrument PXIe-6356 multifunction I/O device, 
connected to a computer, 
is used in experiments to generate the real-time control command 
at a sampling rate of $F_{rf}$ = 1 kHz. 
A LabVIEW Real-Time module is used to execute the program. 
Sensor data acquisition and control command generation 
for the AI control experiments are operated under the same sampling frequency of 1 kHz. 
It has been confirmed that the ON/OFF command lasts at least 1 ms to ensure the actuators to work effectively. 
The available $f_a$ can be derived from $f_a$ = $F_{rf}/N_{sp}$, 
where $N_{sp}$ is the number of sampling points in one period 1/$f_a$. 
The working frequency range of actuators ([0, 500 Hz]) 
imposes a minimum value for $N_{sp}$, i.e. $N_{sp} \ge 2$. 
For a given frequency, $\alpha$ can be deduced from $m/N_{sp}$, $m$ = 1, $\dots$, $N_{sp}$-1. 
The $m$ range ensures a response time of 1 ms for the effective working of the actuators, 
which is adequate as the maximum sampling rate $F_{rf}$ is 1 kHz 
due to the limitation of hardware. 
Thus, the number of possible duty cycles $N_{\alpha}$ for a given $f_a$ is $N_{\alpha}$ = $N_{sp} - 1$ = $F_{rf}/f_a - 1$, 
which increases with $F_{rf}$ and decreases with $f_a$. 
This process is similar to the one used by \citet{Li2017aom} and \citet{Wu2018eif}.

\section{Minijet actuation}
\label{ToC:Minijet actuation}

\subsection{Minijet-produced flow}
\label{ToC:Minijet-produced flow}
%--- Figure ------------------------------------------------------------
\begin{figure}
 	\centering
    \includegraphics[width=0.8\textwidth]{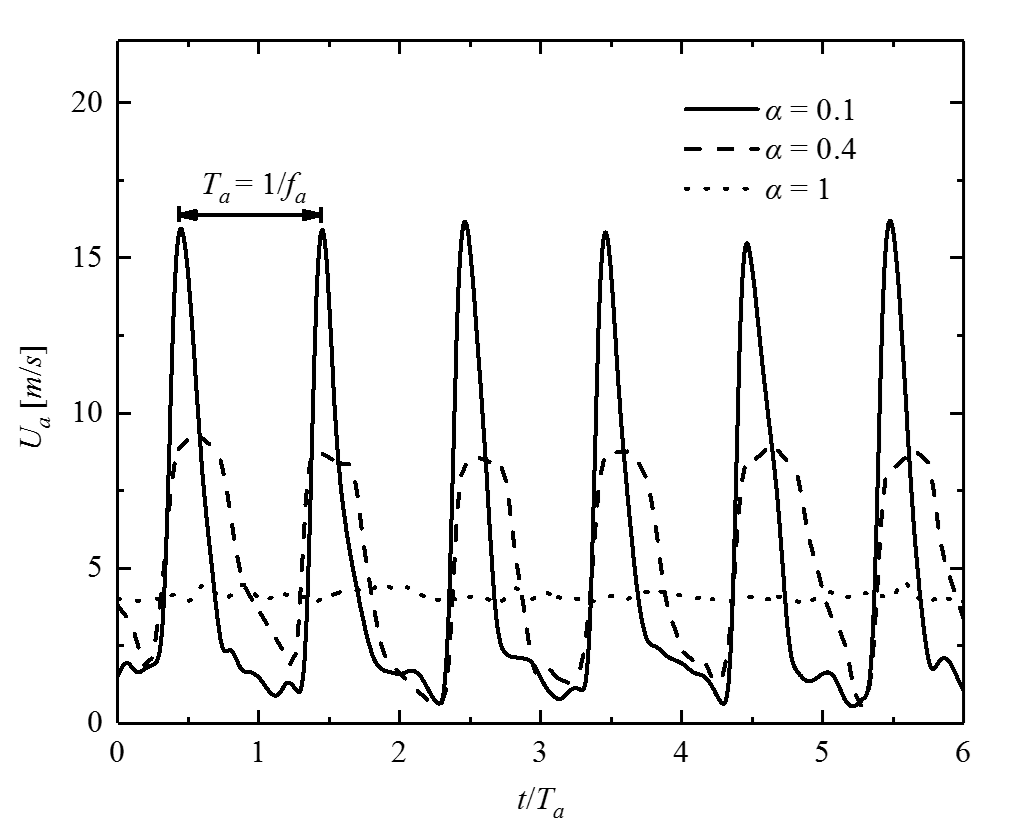}\\
    \caption{ Time histories of minijet injection velocity $U_a$ 
    at duty cycles $\alpha$ = 0.1, 0.4 and 1 (see legend) 
    in the absence of the main jet measured 
    at ($x/D, y/D, z/D$) = (-0.85, -0.35, 0) for $C_m$ = $1.2 $\%, $f_a/f_0$ = $0.5$. $\overline{U_j}$ = 0. }
 	\label{Fig:1mj_signal}
 \end{figure}
%-----------------------------------------------------------------------

It is important to document the flow produced by a minijet 
and the effect of minijets on or the initial condition of main jet. 
This information is crucial for understanding physically the manipulated jet. 
The instantaneous velocity $U_a$ of a single radial minijet is first examined in the absence of main jet.
A hot-wire is placed 17 mm or $x/D$ = -0.85 upstream 
of the main jet exit and 3 mm radially 
from the exit of minijet 1 (figure \ref{Fig:Setup}$c$). 
The hot-wire is oriented normal to the minijet axis—recording the signal $U_a$, 
which changes with $\alpha$ (figure \ref{Fig:1mj_signal}). 
For $\alpha$ = 0.1, $U_a$ displays sharp peaks which are periodic and clearly separated. 
But these peaks are less pronounced at $\alpha$ = 0.4. 
$U_a$ is almost steady at $\alpha$ = 1, 
though showing a small variation, as observed by \citet{Johari1999AIAA}. 
Apparently, a small $\alpha$  produces a large instantaneous velocity, 
implying a large penetration depth into main jet. 

%--- Figure ------------------------------------------------------------
\begin{figure}
    \centering
    \includegraphics[width=0.75\textwidth]{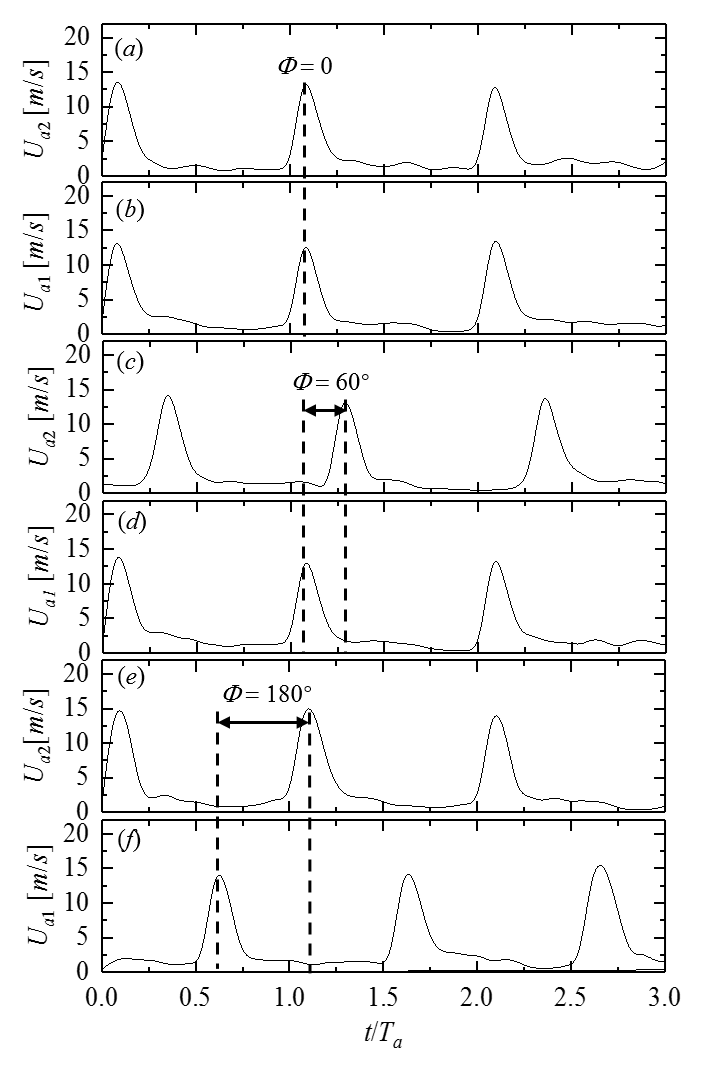}\\
    \caption{Time histories of two minijet injection velocity signals $U_{a1}$, $U_{a2}$ measured simultaneously 
    at ($x/D, y/D, z/D$) = (-0.85, $\pm$ 0.35, 0 ) 
    for $C_m$ = $1.2$\%, $f_a/f_0$ = 0.5 and $\alpha$ = 0.1. 
    There is a phase difference $\Phi$ between two minijets control signals: ($a-b$) $\Phi$ = $0$, ($c-d$) $\Phi$ = $60^\circ$, ($e-f$) $\Phi$ = $180^\circ$. $\overline U_j$ = 0.}
    \label{Fig:2mj_signal}
 \end{figure}
%---------------------------------------------------------------

Consider the simultaneous injection of minijets 1 and 4 (figure \ref{Fig:Setup}$c$) without main jet. 
Two hot-wires are placed perpendicularly to the $x-y$ plane at $x/D$ = -0.85 and 
3 mm from each of the corresponding measured minijet exit. 
The two minijets are injected with a phase shift $\Phi$, 
which may be varied by changing the phase shift 
between the two square wave signals of input voltages. 
At $\Phi$ = 0$^\circ$, the $U_{a1}$ signal exhibits a very sharp peak value, 
with a magnitude of close to 0 at the off-state of the minijet 
and about 13 at the on-state (figure \ref{Fig:2mj_signal}$a$). 
Note that, even after the electromagnetic valve is closed, 
there may be some fluid injecting into main jet \citep{Sailor1999ijhff}. 
A similar observation can be made for $\Phi$ = 60$^\circ$ and 180$^\circ$ (figure \ref{Fig:2mj_signal}$b$, $c$). 
The characteristics of $U_{a2}$ resemble those of $U_{a1}$, 
regardless of the $\Phi$ value. 
It may be inferred that each of the minijets does not 
depend on $\Phi$ and is rather independent of each other.

\subsection{Penetration depth and minijet number}
\label{ToC:Penetration depth and minijet number}

%--- Figure ------------------------------------------------------------
\begin{figure}
 	\centering
    \includegraphics[width=1\textwidth]{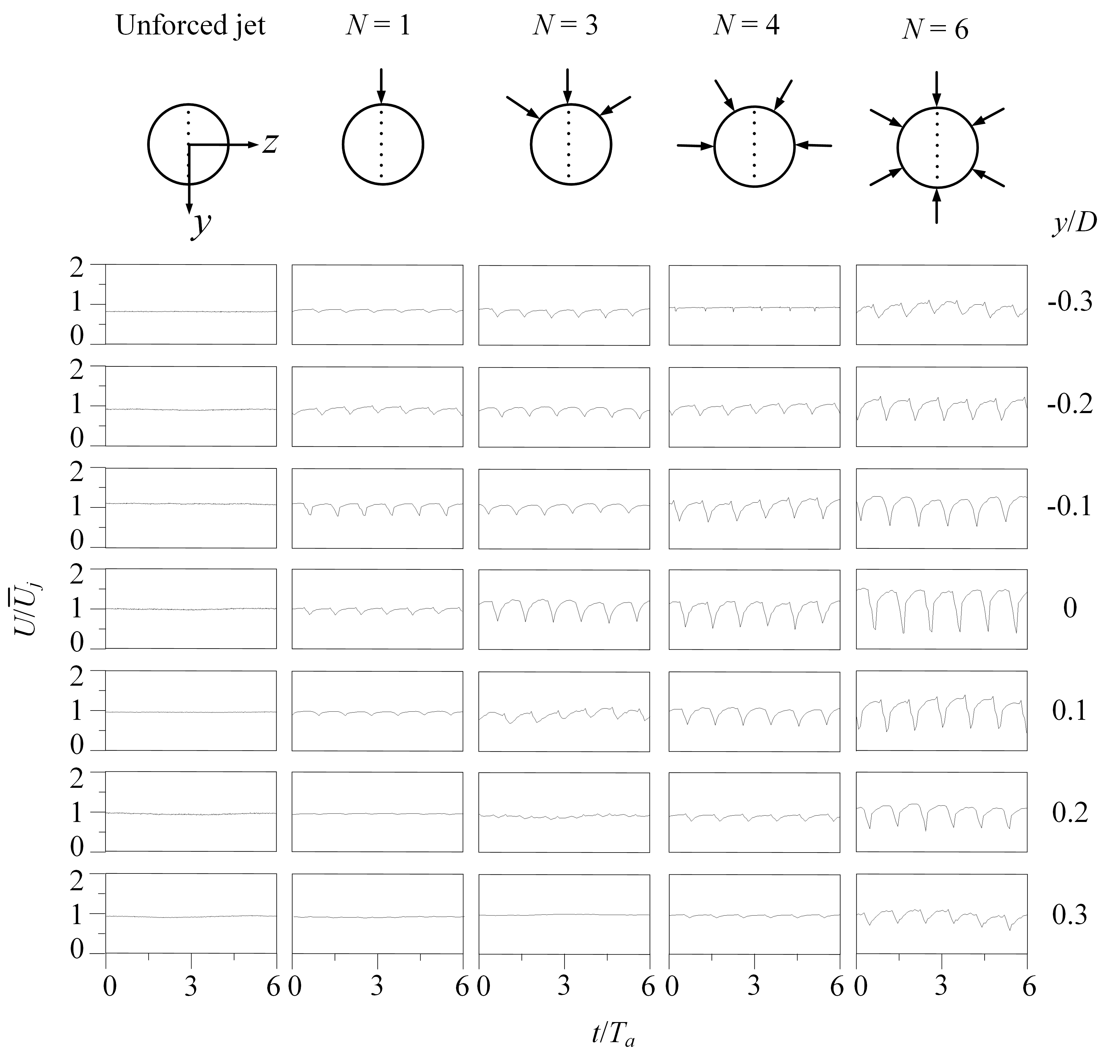}\\
    \caption{Typical hot-wire signals of  instantaneous streamwise velocity $U/\overline{U_j}$ along $x-y$plane 
    at $x/D$ = $0.05$ for minijet number $N$ = 1, 3, 4 and 6 
    ($f_a/f_0$ = $0.5$, $\alpha$ = $0.15$, $C_m$ =$1.2\%$). 
    The same scale is applied for all signals. 
    The dots within the circle represent the hot-wire measurement points.
    }
 	\label{Fig:multi_mj_signal}
 \end{figure}
%-----------------------------------------------------------------------

The penetration depth of control jets may have a pronounced impact upon jet mixing \citep{Davis1982AIAA}. 
Thus, its influence on main jet is examined for 
various minijet numbers and configurations with given $\overline U_j$ and control parameters, 
i.e. $\alpha$, $f_a/f_0$ and mass flow ratio $C_m$ = $m_{mini}/m_j$, 
where $m_{mini}$ and $m_j$ are the mass flow rates of a single minijet and main jet, respectively. 
The minijet penetration depth could be approximately estimated from the $U$ signals along the $y$ direction, 
measured at $x/D$ = 0.05 using a hot-wire placed perpendicularly to the $x-y$ plane, 
as shown in figure \ref{Fig:multi_mj_signal}, 
where the scale of the abscissa or ordinate is made the same for all cases to facilitate comparison. 
The $U$ signals are essentially constant throughout the range of $y/D$ $\in $ [-0.3, 0.3] for the unforced jet. 
The periodic fluctuations of $U$ appear at $y/D$ = - 0.3 for one minijet injection ($N$ = 1), 
and its magnitude grows first and then retreats with increasing $y/D$. 
The fluctuations remain discernible at $y/D$ = - 0.1. 
Note that the minijet is issued along the $y$ direction. 
Beyond $y/D$ = - 0.1, 
the velocity fluctuations are negligibly small and in fact comparable to that in the unforced jet. 
These observations indicate that the minijet has reached a penetration depth of $y/D$ = - 0.1. 
With three adjacent minijets on ($N$ = 3), 
the velocity fluctuations are appreciably larger in magnitude 
than their counterparts of $N$ = 1, and the maximum amplitude 
is shifted to a deeper position, i.e. from $y/D$ = - 0.1 
at $N$ = 1 to $y/D$ = 0 at $N$ = 3. 
The fluctuations are now discernible at $y/D$ = 0.2, 
indicating an increased penetration depth, 
though the minijets clearly have not impinged on the wall opposite to the injecting minijets. 
With $N$ increasing to 4, 
the maximum magnitude of the velocity fluctuations is appreciably larger than that of $N$ = 3, 
and again occurs at the centre ($y/D$ = 0)
where all minijets contribute to an increase in the velocity fluctuations. 
Furthermore, the fluctuations are now even discernible at $y/D$ = 0.3. 
It is worth pointing out that we did not move the hot-wire closer to the wall because of its high fragility; 
therefore, we could not tell whether the minijets have penetrated through the main jet in this case. 
At $N$ = 6, the velocity fluctuations display symmetry about the centre, 
the maximum magnitude exceeding all other cases and taking place at the centre. 

\subsection{Power spectral density function and minijet number}
\label{Power spectral density function and minijet number}

%--- Figure ------------------------------------------------------------
\begin{figure}
 	\centering
    \includegraphics[width=1\textwidth]{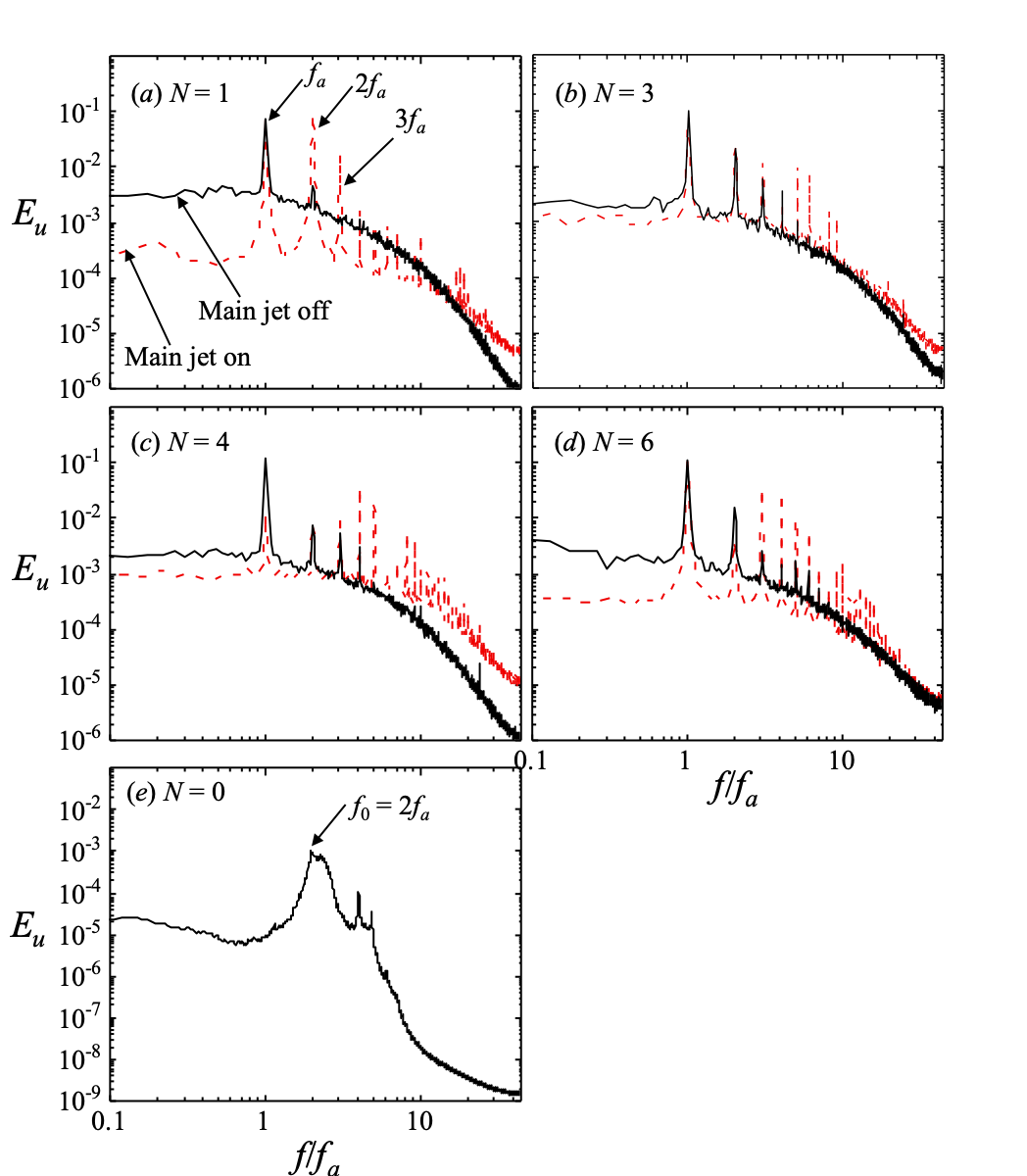}\\
    \caption{Power spectral density function $E_u$ of hot-wire signals $u$ 
    measured at($x/D, y/D, z/D$) = (0.05, 0, 0) with and without main jet: ($a$) $N$ = 1, ($b$)  $N$ = 3, ($c$)  $N$ = 4, ($d$)  $N$ = 6, ($e$)  $N$ = 0,($x/D, y/D, z/D$) = (3, 0, 0).}
 	\label{Fig:exit_PSD}
 \end{figure}
%-----------------------------------------------------------------------

Figure \ref{Fig:exit_PSD} compares $E_u$ measured on the centreline 
at $x/D$= 0.05 with and without main jet operated, 
where the log-log scale is used to emphasize the low-frequency components.
This function $E_u$ yields  $\overline{u^2}$ = $\int_{0}^{\infty}E_udf$, 
where $f$ is frequency. $E_u$ (Figure \ref{Fig:exit_PSD}$e$) 
measured in the unforced jet shows a pronounced peak at $f_0$, 
indicating clearly the occurrence of the preferred mode structure. 
When the minijets as well as main jet are operated, 
$E_u$ exhibits more pronounced peaks at $f/fa$ = 1 and its harmonics. 
These observation results from the interaction between main jet and minijet, 
referred to as the parametric resonance by \citet{Huang1999pof}. 
Evidently, the unsteady injection produces the periodic structures 
upstream of the nozzle exit, as noted by \citet{Zhou2012AIAA}. 
With increasing $N$, 
the peaks become more pronounced and occur at more harmonics, 
echoing the enhanced periodic structures (figure \ref{Fig:multi_mj_signal}) 
and hence the enhanced excitation of the shear layer. 
The predominant frequencies do not vary with $N$ though. 
Note that $E_u$ is normalized by $\overline{u^2}$ 
so that its integration over the entire frequency range is always 
equal to unity. 
As a result, $E_u$ drops appreciably over the low frequency range.

\subsection{Fluctuating velocity and minijet number}
\label{Fluctuating velocity and minijet number}

%--- Figure ------------------------------------------------------------
\begin{figure}
 	\centering
    \includegraphics[width=1\textwidth]{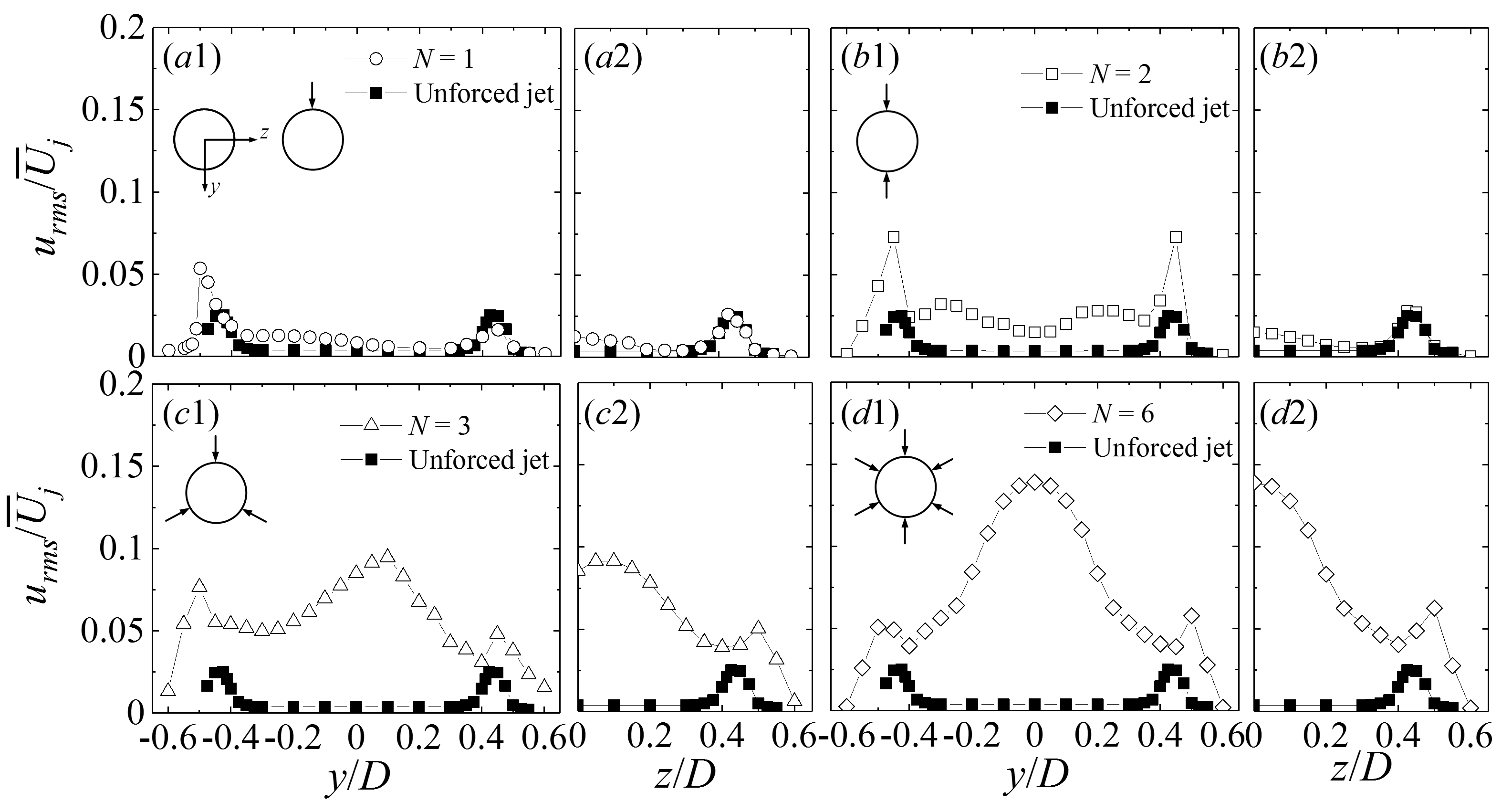}\\
    \caption{Radial distributions of fluctuating velocity $u_{rms}$/$\overline{U_j}$ measured 
    at $x/D$ = 0.05 ($f_e$/$f_0$ = $0.5$, $C_m = 1.2\%$) depend on minijet number: ($a1-d1$) along the $y$ axis, ($a2-d2$) along the $z$ axis.
    }
 	\label{Fig:exit_urms}
 \end{figure}
%-----------------------------------------------------------------------

The number and configuration of minijets may profoundly affect the main jet issuing from the nozzle, 
as the case of passive delta tabs \citep{zaman1994pof}. 
This effect plays an important role in the downstream evolution of flow.
As such, the radial profiles of the hot-wire 
measured root mean square ($rms$) velocity $u_{rms}$ at $x/D$ = 0.05 
are examined in the manipulated jets, along the $y$ and $z$ axes, 
respectively, for $N$ = 1, 2, 3, 6. 
The data of the unforced jet are also presented for the purpose of comparison. 
Given the symmetrically arranged minijets ($N$ = 2, 6) about the $x-z$ plane (figure \ref{Fig:exit_urms}$b1$,$d1$), 
the $u_{rms}$ distributions along the $y$-axis exhibit reasonable symmetry. 
The $u_{rms}$ displays a pronounced peak about $y/D$ = $\pm 0.45$, 
where the shear layer is expected, in the injection or $x-y$ plane 
for $N$ = 2 (figure \ref{Fig:exit_urms}$b1$) 
but remains unchanged in the orthogonal or $x-z$ plane (figure \ref{Fig:exit_urms}$b2$), 
indicating that the shear layer between the two minijets 
is essentially undisturbed. 
Being symmetrical about the $z$ axis, the $u_{rms}$ distributions 
are given only for $ z/D \ge$ 0 in figure \ref{Fig:exit_urms}($a2-d2$). 
A broad bump is evident at $y/D \approx$ 0.2 for $N$ = 2 (figure \ref{Fig:exit_urms}$b1$). 
The flow structure induced by an unsteady injecting minijet 
is similar to a pulsed jet in cross flow, 
which forms a series of periodical vortex rings \citep{closkey2002jfm}. 
It seems that these minijet-produced periodic vortices may occur most likely at $y/D \approx$ 0.2, accounting for the broad bump. 
For $N$ = 3 and 6, this bump moves to near the centre, 
with a significantly increased magnitude (figure \ref{Fig:exit_urms}$c1$, $d1$). 
Two factors may be responsible for this increase. 
Firstly, as the separation angle $\theta$ decreases from 180$^\circ$ to 120$^\circ$ and then 60$^\circ$, 
two neighbouring minijets become close and their induced unsteady flows interact more and more intensely. 
\citet{zaman1994pof} noted that, 
as the neighboring delta tabs approach each other, 
streamwise vortices interact more vigorously, 
resulting in the jet core fluid ejection. 
Secondly, as demonstrated in figure \ref{Fig:multi_mj_signal}, 
every minijet generates a velocity fluctuation at the centre. 
For $N$ = 6, $\theta$ is smallest and all six minijets 
contribute to flow perturbations, 
thus producing the most pronounced bump at $y/D$ = 0.

\section{Artificial intelligence control system}
\label{ToC:AIControl_system}

\subsection{Artificial intelligence control system}
\label{ToC:Artificial intelligence control system}

%--- Figure ------------------------------------------------------------
\begin{figure}
\begin{center}
\includegraphics[width=0.7\textwidth]{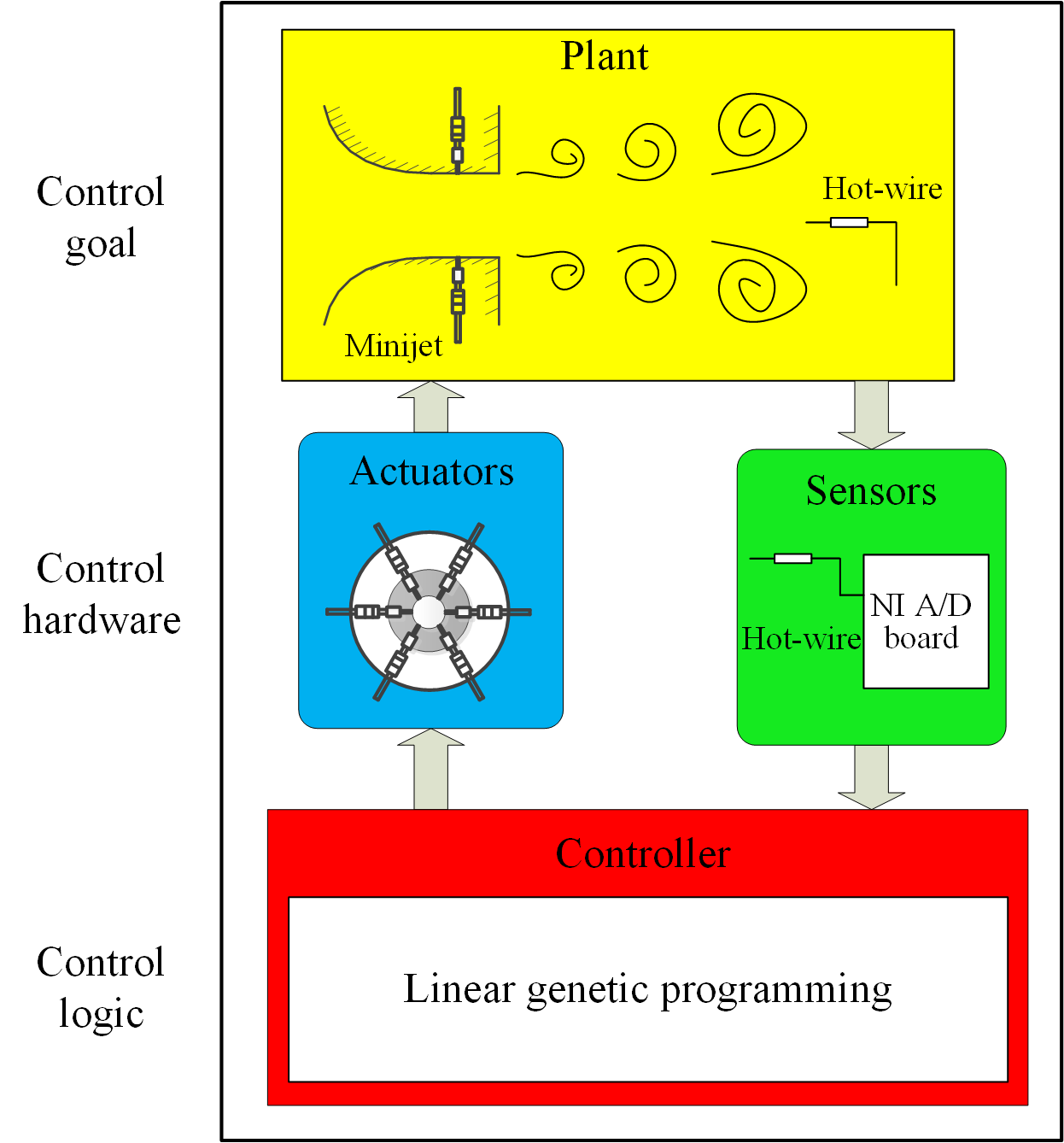} \\
\end{center}
\caption{Principle sketch of the artificial intelligence control
which consists of a plant (yellow), sensors (green), actuators (blue)
and a controller (red) that includes linear genetic programming (LGP) algorithm or other machine learning methods.
}
\label{fig:AIC_box}
\end{figure}
%--- Figure ------------------------------------------------------------

Artificial intelligence methods allow us to 
explore the rich universe of nonlinear actuation mechanisms opened 
by independent spatially distributed actuators. 
Hence, we see the actuation and sensing hardware and control logic as intimately interwoven. 
The AI control system is sketched in figure \ref{fig:AIC_box}. Generally, a control system facilitates a control goal 
for a plant by control hardware and a control logic/controller. 
The control hardware includes sensors and actuators as discussed in \S~\ref{ToC:Exp_details}. 
This hardware monitors the plant output (velocity signals) and 
executes instructions from the controller. 
The open-loop arrangement is shown in figure \ref{Fig:Setup}$a$ 
for calculating the cost value $J$ = $\overline U_{5D}/\overline U_j$. 
A minimized cost $J$ corresponds to the maximized decay rate $K$ = $1-J$ of jet centreline mean velocity, 
which is an indicator of the mixing efficacy of a jet \citep{ARUN2018jfm}.

\subsection{Control optimization using linear genetic programming}
\label{ToC:Control optimization}

The six-dimensional vector $\boldsymbol{b} = [b_1,b_2,\ldots,b_6]^{\dagger}$ comprises all actuation commands.
The $i$th minijet is `ON'
if the actuation command $b_i$ is positive
and is `OFF' otherwise. 
In the sequel, we assume $b_i$ = 1 for `ON', and $b_i$ = 0 for `OFF'.
Following \citet{Wu2018eif},
we search for a control law including sensor-feedback with hot-wire signals $\boldsymbol{s}$, 
multi-frequency open-loop forcing with harmonic functions contained 
in $\boldsymbol{h}=[h_1,h_2\ldots,h_{6}]^{\dagger}$. 
Here, $h_i= \sin \left( \omega_a t - \phi_i \right)$, 
$i=1,2,\ldots,6$, where $t$ is time, 
$\omega_a$ is a reference frequency to be determined in \S~\ref{ToC:Representative reference actuations} and $\phi_i$ is the phase. 
Then,

\begin{equation}
\boldsymbol{b} = \boldsymbol{K} (\boldsymbol{s}, \boldsymbol{h}),
\label{Eqn:ControlLaw}
\end{equation}
where the vector function $\boldsymbol{K} = [K_1\ldots,K_6]$ comprises the actuation laws for each minijet. 
The time-averaged duty cycle of the $i$th minijet is determined 
by the control law $\boldsymbol{K}$ and the arguments, 
i.e. sensor signals and harmonic functions. 
For open-loop forcing $\boldsymbol{b} = \boldsymbol{K}(\boldsymbol{h})$, 
the duty cycle of the $i$th minijet becomes the sensor-independent time-average 
of the actuation command $\overline K_i (h)$. 
Thus, helical forcing may have a particularly simple representation, e.g. $b_i = h_i$. 
In general, only two harmonic functions, typically $\sin \omega_a t$ and 
$\cos \omega_a t$, 
are sufficient for harmonic functions with arbitrary phases. 
Following \citet{Paschereit1995jfm} and others, 
we add the cosine and sine components of $\omega_a$/2 and $\omega_a$/4, 
yielding a ten-dimensional vector $h = [h_1, h_2\ldots, h_{10}]^{\dagger}$. 
The nonlinear function $\boldsymbol{K}$ can create arbitrary higher harmonics, 
arbitrary phase relationships between $\omega_a$, $\omega_a$/2, $\omega_a$/4, 
and higher harmonics, 
e.g. $1 - 2h^{10}$ = cos(10$\omega_a t$), as well as arbitrary sum and difference frequencies. 
The control optimization searches for a law of form \ref{Eqn:ControlLaw} that minimizes the cost,

\begin{equation}
\boldsymbol{K}^{\star} = \arg\min\limits_{\boldsymbol{K}} J \left [\boldsymbol{K} \right].
\label{Eqn:RegressionProblem}
\end{equation}
The regression problem implies a search for a mapping from multiple inputs to a multiple-output signal. 
Even in case of a linear function this implies the optimization of a large number of parameters. 
We employ the powerful linear genetic programming (LGP) 
as a regression solver and take the same parameters for 
the control law representation and for the genetic operations as \citet{Wu2018eif}. 
The first generation of LGP, $n$ = 1, contains $N_i$ = 100 random control laws, also called individuals. 
Each individual `$i$' is experimentally tested for 5 seconds 
to yield the measured cost $J_i^n$, where superscript `$n$' denotes the generation number. 
Subsequent generations are produced from the previous ones with genetic operations (elitism, crossover, mutation and replication) and tested analogously. 
Elitism pass directly the top-ranking individuals to next generation. 
Replication copies a stochastically selected number of individuals into next generation, 
which acts to preserve some well performing individuals. 
Crossover involves two selected individuals and then produces two individuals, 
with part of their elements exchanged. 
This operation tends to generate better individuals by exploitation. 
For the mutation operation, the instructions of a selected individual are randomly changed. 
Both crossover and mutation serve to explore potentially new 
and better minima of $J$. 
After the in situ performance measurements, 
the individuals are re-numbered in order of performance, $J_1^n \le J_2^n \le \ldots \le J_{N_i}^n$,
where subscript $i$ represents the individual index and $N_i$ and $n$ denote the size and number of generations. 
We have noted in trial tests that all the winning individuals always involve every actuator. 
Therefore, when generating the 100 individuals of the first generation, 
we exclude the possibility of permanently inactive actuator to accelerate the learning process, 
that is, as a plant-specific rule, 
we discard and replace any individual for testing if one or more actuators are not active. 

It is worth mentioning that the present jet control is formulated as a model-free regression problem: 
determine the law which minimizes the given cost function. 
The considered search space of control laws significantly extends hitherto considered actuations. 
First, general multiple-input actuation is allowed without any imposed symmetry constraints. 
Thus, actuations with arbitrary combinations of minijets thereof can be realized. 
Second, the search space includes broadband multi-frequency actuation. 
Third, nonlinear sensor-feedback is included, 
which is made by nonlinear operations with the sensor signal $\boldsymbol{s}$, e.g. $\boldsymbol{b}$ = $log_{10}$($\boldsymbol{s}$) \citep{Wu2018eif}.
However, this feature is not found improving appreciably the control performance and is therefore removed eventually in the learning process. 
Fourth, the control law may include nonlinear combinations of multi-frequency forcing and sensor-feedback. 
The key enabler for the control optimization in this search space is genetic programming as powerful regression solver. 
Genetic programming may be considered as an example for the many powerful regression solvers of AI.

\subsection{Parameters and control landscape}
\label{Toc:Parameters and control landscape}

\begin{table}
	\begin{center}
	\def~{\hphantom{0}}
	 \begin{tabular}{cc}
		%\toprule  % 顶部线
		Parameters & Value \\[3pt]
		
		%\midrule  % 中部线
		Individuals & $N_i$ = 100 \\
		Tournament size & $N_t$ = 7 \\
		Elitism & $N_e$ = 1 \\
		Crossover & $P_c$ = 70$\%$\\
		Mutation & $P_m$ = 20$\%$ \\
		Replication & $P_r$ = 10$\%$ \\
		Min. instruction number	& 10 \\
		Max. instruction number & 50 \\
		Operations & $N_i$ = $+, -, \times, \div, \sin, \cos, \tanh, \log, g^2$ \\
		Number of constants & $N_c$ = 3 \\
		 Constant range & [-1, 1] \\
		%\bottomrule  % 底部线
	\end{tabular}
\caption{Linear genetic programming parameters employed for experiments. 
The symbol $g$ indicates an input argument.
}
\label{Table:Parameter}
\end{center}
\end{table}

The LGP parameters for this study are displayed in Table \ref{Table:Parameter}. 
These values are adopted from a previous MLC jet mixing study in the same facility with a single minijet \citep{Wu2018eif}. 
The parameters are identical or close to the ones employed 
in numerous experimental studies as summarized by \citet{Duriez2016mlc} and \citet{Noack2019fssic}. 
Elitism is set to $N_e$ = 1, 
i.e. the best individual of a generation is copied to the next one. 
The replication, crossover and mutation probabilities are $10\%, 70\%$ and $20\%$, respectively. 
The individuals on which these genetic operations are performed 
come from a tournament selection of size $N_t$ = 7. 
The instruction number varies from 10 to 50. 
The operations comprise $+, -, \times, \div, \sin, \cos, \tanh, \log_{10}$ and $g^2$, 
where $g$ is the input argument.
The operations $\div$ and $\log_{10}$ are protected to prevent an undefined expression with a vanishing argument; 
for example,  $\log_{10}(g)$ is modified to $\log_{10}(\vert g\vert)$. 
In addition, LGP uses three random constants in the range [-1, 1].

The evolution of control laws is depicted with a proximity map following \citet{Duriez2016mlc}. 
The main idea is that the considered ensemble of 
$\boldsymbol{K}_i(\boldsymbol{h})$ is represented as points 
in a two-dimensional feature plane $\boldsymbol{\gamma}_i$ = ($\gamma_{i,1}, \gamma_{i,2}$), where $i = 1,2\ldots, N_i \times n$, 
so that the difference between the control laws is optimally indicated by distance between feature vectors. 
The key is the definition of a metric $D_{ij}$ 
between the control laws $\boldsymbol{K}_i(\boldsymbol{h})$ and $\boldsymbol{K}_j(\boldsymbol{h})$. 
For the considered open-loop actuation, 
this metric is the root-mean-square averaged Euclidean difference 
between the actuation command vectors 
accounting for a potential time-delay, given by

\begin{equation}
M_{ij} = \min\limits_{\tau \in [0, T_a]} \sqrt{\Vert \boldsymbol{K}_i(\boldsymbol{h}(t))-\boldsymbol{K}_j(\boldsymbol{h}(t-\tau))\Vert^2}.
\label{Eqn:Metric}
\end{equation}
In the employed metric, 
we incorporate also the control performance $J_i$ by a penalization term, i.e.

\begin{equation}
D_{ij} = M_{ij}+ \beta \vert J_i-J_j\vert
\label{Eqn:Distance}
\end{equation}
The parameter $\beta$ is chosen so that the 
maximum actuation distance $M_{ij}$ is equal to the maximum difference in the performance terms:

\begin{equation}
\max\limits_{i,j = 1,2\ldots N_i \times n} M_{ij}=\beta\max\limits_{i,j = 1,2\ldots N_i \times n} \vert J_i-J_j\vert.
\label{Eqn:Max distance}
\end{equation}
Given the resulting configuration matrix $\boldsymbol{D}$ = ($D_{ij}$) ($i, j = 1, 2\ldots, N_i \times n$), 
classical multi-dimensional scaling \citep{Cox2001MS} uniquely 
determines feature vectors $\boldsymbol{\gamma}_i$, $i = 1, 2\ldots, N_i \times n$, 
so that the distances are optimally preserved:

\begin{equation}
\sum_{i=1}^{N_i \times n} \sum_{j=1}^{N_i \times n}(\Vert \boldsymbol{\gamma}_i-\boldsymbol{\gamma}_j\Vert-D_{ij})^2=min.
\label{Eqn:min distance}
\end{equation}
The translational degree of freedom is removed by centering 
the feature vectors $\sum_{i=1}^{N_i \times n} \boldsymbol{\gamma}_i=0$. 
The feature vectors are sorted and rotated so that the first coordinate has the largest variance, 
the second coordinate the second largest, etc. 
The coordinates are indeterminate by a sign (mirroring), 
like POD modes and their amplitudes.

Finally, a control landscape $J(\boldsymbol{\gamma})$ is interpolated 
from the three-dimensional data points 
($\gamma_{i,1}, \gamma_{i,2}, J_i$), $i = 1, 2\ldots, N_i \times n$. 
The two-dimensional feature vectors $\boldsymbol{\gamma}_i$ are connected by an unstructured grid 
from a \citet{delaunay1934Sur} triangulation. 
This triangulation guarantees that the mesh triangles are optimally equilateral. 
The $J$-values in each mesh triangle $i_1, i_2, i_3$ $\in$ 
$\{ 1,2\ldots, N_i \times n \}$ are interpolated from the known values 
at the vertices $J_{i_1} , J_{i_2} , J_{i_3}$. 
These control landscapes have been employed in several 
AI-based control schemes \citep{Kaiser2017tcfd}. 
They indicate the complexity of the actuation response 
and the learning progress of AI-based control. 
Often, the feature coordinates can be linked with 
the physical properties of actuation a posteriori, 
thus providing additional insights.

\section{Outcome of the AI control }
\label{ToC:Learningcontrollaw}

\subsection{Representative reference actuations}
\label{ToC:Representative reference actuations}

A few well-known reference forcings are firstly presented to facilitate 
the understanding of the AI learning process and 
highlight the uniqueness of this method. 
In our earlier studies, turbulent jet mixing has been optimized 
for the same cost function and experimental conditions. 
For single unsteady minijet forcing, the optimal $f_a$ is found 
to be 67 Hz \citep{Wu2018eif}, 0.5$f_0$, and the optimal $C_m$ is 1.2\% 
based on a dual-input-and-one-output closed-loop control technique \citep{Wu2018AIAA}. 
As such, we choose the same $f_a$ or $\omega_a$ = 2$\pi f_a$ and $C_m$ = 1.2\% for every minijet. 
With $C_m$ fixed for each minijet, 
the overall mass flow of injected fluid in one actuation period $T_a$ is the same for all actuations, 
that is, the input/actuation energy is the same, 
irrespective of control modes or laws. 
Consider three reference forcings (e.g. \citet{Hilgers2001fdr}, \citet{Yang2016jfm}, \citet{Yang2017phd}), viz.

\begin{subequations}
\begin{eqnarray}
\hbox{\emph{axisymmetric forcing}}       && b_i = h_1 - \alpha_a , \quad i=1,2,\ldots,6;
\label{Eqn:AxisymmetricForcing}   \\
\hbox{\emph{helical forcing}}            && b_i = h_i - \alpha_h, \quad i=1,2,\ldots,6;
\label{Eqn:HelicalForcing}        \\
\hbox{\emph{flapping forcing}} && b_1=b_2=b_3= h_1 -\alpha_f,  b_4=b_5=b_6 =h_4-\alpha_f .
\label{Eqn:FlappingForcing}
\end{eqnarray}
\label{Eqn:SymmetricForcing}
\end{subequations}%
The constants $\alpha_a$, $\alpha_h$ and $\alpha_f$ 
correspond to the duty cycles and have been optimized 
with respect to the cost. 
As mentioned before, actuation is performed only when $b_i > 0$. 
The cost functions are found to be $J_a$ = 0.665, $J_h$ = 0.568 and $J_f$ = 0.423 
for the optimized axisymmetric, helical and flapping forcings (Table \ref{Table:benchmark}), respectively, 
based on the conventional open-loop control, 
which provide the benchmarks for the AI control performance to be discussed below.

\begin{table}
    \centering
    \begin{tabular}{ccc}
       % \toprule  % 顶部线
        Mode &Benchmark forcing & AI control\\[3pt]        
        %\midrule  % 中部线
        Unforced jet & $J_u$ = 0.974 & \\
        Axisymmetric forcing & $J_a$ = 0.665 & $J_1^1$ = 0.626\\
        Helical forcing & $J_h$ = 0.568 & $J_1^2$ = 0.555\\
        Flapping forcing & $J_f$ = 0.423 &$J_1^5$ = 0.419\\
        Combined forcing &  & $J_1^{11}$ = 0.305\\    
        %\bottomrule  % 底部线
    \end{tabular}
\caption{Cost function $J$ for different actuations at $Re_D$ = 8000. 
}
\label{Table:benchmark}
\end{table}

\subsection{Learning process of AI control}
\label{ToC:Learning process of AI contro}

In the initial stage of the learning process, 
we included a feedback signal $\boldsymbol{s}$($t$) = [$u_{3D}, u_{5D}$] 
as one input (eq. \ref{Eqn:ControlLaw} for the AI system, 
where $u_{3D}$ and $u_{5D}$ are the fluctuating velocity signals 
measured at $x/D$ = 3 and 5, respectively. 
It is found from dozens of experiments that the search for 
the optimal solution benefits neither from the subharmonic components 
of $h_7,\ldots, h_{10}$ nor from the feedback signals. 
Therefore, the AI control laws \ref{Eqn:ControlLaw} may be cast 
in the periodic open-loop form, $b_i = K_i (h_1, h_2\ldots, h_6)$, 
$i = 1, 2,\ldots, 6$. 
Hence, we restrict our following discussion to this open-loop actuation. 

%--- Figure ------------------------------------------------------------
\begin{figure}
    \centering
    \includegraphics[width=0.9\textwidth]{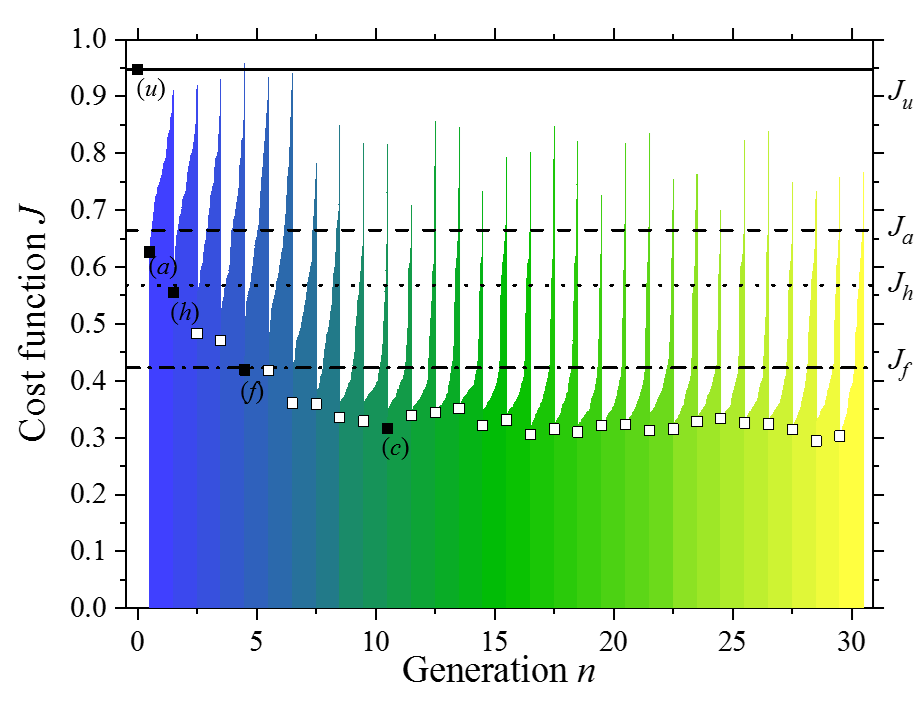}\\
    \caption{Learning curve ($3000$ individuals) of AI control for 
    ($u$) unforced jet, 
    ($a$) axisymmetric forcing, 
    ($h$) helical forcing, 
    ($f$) flapping forcing and 
    ($c$) combined forcings. 
    $J_u$, $J_a$, $J_h$ and $J_f$ are costs corresponding to the benchmarks of unforced, 
    open-loop axisymmetric, helical and flapping forcings, respectively (see table \ref{Table:benchmark}).}
    \label{Fig:LearningCurve}
 \end{figure}
%-----------------------------------------------------------------------

The learning curve of the AI control is presented in figure \ref{Fig:LearningCurve}, 
where the square symbol marks the first and best individual of each generation with $N_i$ = 100 control laws. 
The remaining costs grow monotonously with their indices, 
and the 100 individuals of each generation form a color bar. 
The square symbol curve unveils the best performance from generation $n$ = 1 to 30. 
The best individual of the first generation or stage 1 is characterized by an axisymmetric control law 
(see Eq. \ref{Eqn:generation_1} in Appendix \ref{ToC:Appendix}). 

This law is equivalent to \ref{Eqn:AxisymmetricForcing} except for a time shift, reflected by $4/6\pi$. 
The performance $J_1^1$ = 0.626 (see figure \ref{Fig:LearningCurve} 
and table \ref{Table:benchmark}) is slightly better, about 5.8\% lower, 
than the benchmark of axisymmetric forcing \ref{Eqn:AxisymmetricForcing}, 
though much higher than that ($J_u = 0.947$ or $K \approx 0.05$) of the unforced jet. 
Note that the centreline mean velocity for calculating $K$ or $J_a$ 
is measured over a duration of 60 seconds in the benchmark experiments, 
but only 5 seconds for estimating $J_i$ as the measured $J_i$ 
is used to evaluate control laws and does not need to be very accurate 
in the learning process of AI control. 
An accurate long-time evaluation of $J$ is performed only 
in the last generation $n$ = 30. 
This difference in evaluating the cost function could account 
for the deviation between $J_1^1$ and $J_a$. 
However, the AI control or specifically genetic programming breeds several copies of the winning individual, 
covering all possible combinations of the control parameters, 
and then takes the best performing one. 
This process differs in essence from the searching process of the conventional open loop control 
which optimizes one control parameter first and then moves to next with the first parameter fixed. 
The advantage of the former over the latter is obvious. 
Therefore, different search strategies cannot be excluded from the mechanisms behind the deviation, 
which will be further substantiated by the fact that all the 
best cost functions of the different stages 
in the learning curve are less, 
albeit slightly, than their corresponding benchmarks produced 
from the conventional open-loop control.

Stage 2 starts with the second generation when the AI control 
discovers a better performing helical forcing (Eq. \ref{Eqn:generation_2} in Appendix \ref{ToC:Appendix}). 
This forcing differs in form from \ref{Eqn:HelicalForcing}, 
but clearly shows a uniformly traveling wave in the azimuthal direction 
(to be demonstrated in \S~\ref{ToC: Representative control laws and flow responses}), 
its cost $J_1^2$ being again slightly lower than $J_h$ (Table \ref{Table:benchmark}). 
Helical forcing reduces $J$ further as found from the numerical simulation study of a similar jet mixing optimization \citep{Hilgers2001fdr}. 
Local spatial stability analysis indicates that, unlike axisymmetric forcing, helical perturbations are spatially amplified downstream of the potential core \citep{Garnaud2013jfm}.

Flapping forcing takes place in stage 3, 
starting from the fifth generation. The law (Eq. \ref{Eqn:generation_5} in Appendix \ref{ToC:Appendix}) is similar to \ref{Eqn:FlappingForcing} but incorporates an asymmetry. 
An optimized asymmetry yields a reproducibly better mixing, again $J_1^5 \le J_f$ (table \ref{Table:benchmark}). 

The eleventh generation marks the emergence of stage 4. 
AI control discovers a very sophisticated control law. 
See Eq. \ref{Eqn:generation_11} in Appendix \ref{ToC:Appendix}. 
This forcing significantly outperforms the flapping forcing 
found in generation 5, the corresponding $J_1^{11}$ plunging to 0.305, 
a drop of 27\% compared with the smallest $J_1^5$ in generation 5 
and less than 1/3 of the unforced jet. 
The actuation mechanism does not change any more 
in following generations with little variation in costs, 
pointing to the convergence of the AI learning process. 
It is worth highlighting that this actuation mechanism is reproducible, 
that is, approximately the same converged cost has been observed 
in all experiments, notwithstanding a change in the initial parameters of the first generation. 
However, not all AI learning curves go through the stages of axisymmetric, helical and flapping forcings; some AI experiments may find only two of the three stages in the learning process.

\subsection{Representative control laws and flow responses}
\label{ToC: Representative control laws and flow responses}

%--- Figure ------------------------------------------------------------
\begin{figure}
    \centering
    \includegraphics[width=0.95\textwidth]{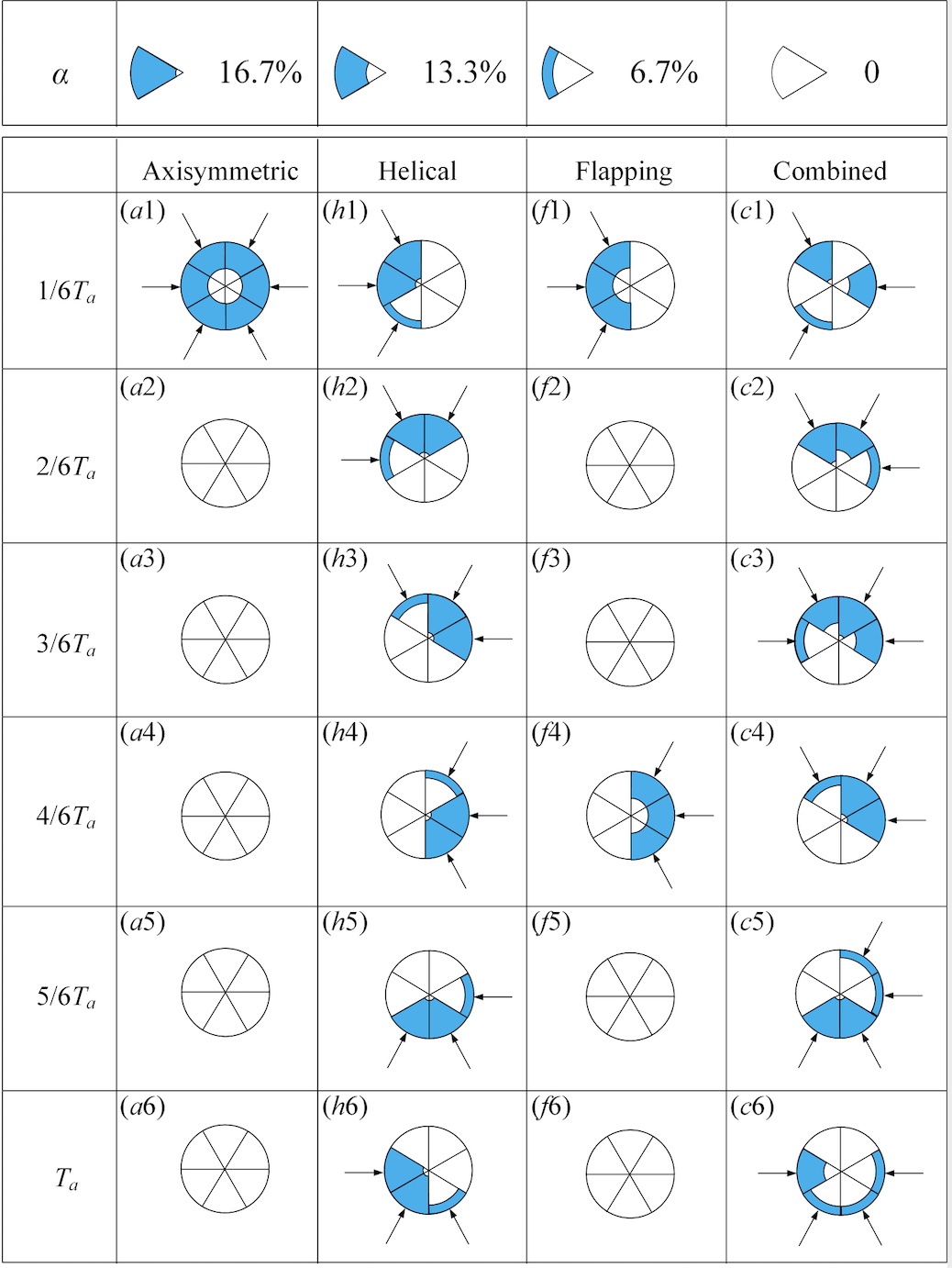}\\
    \caption{Actuation associated with the best individuals of generation $n = 1, 2, 5$ and $11$ of figure \ref{Fig:LearningCurve}. 
    From top to bottom: six instances associated with $\phi$ = $T_a/6, 2T_a/6, 3T_a/6\ldots, T_a$ in one excitation period. 
    Each of the six sectors of the circles corresponds to one minijet.
    }
    \label{Fig:Actuation}
 \end{figure}
%-----------------------------------------------------------------------

%--- Figure ------------------------------------------------------------
\begin{figure}
    \centering
    \includegraphics[width=0.9\textwidth]{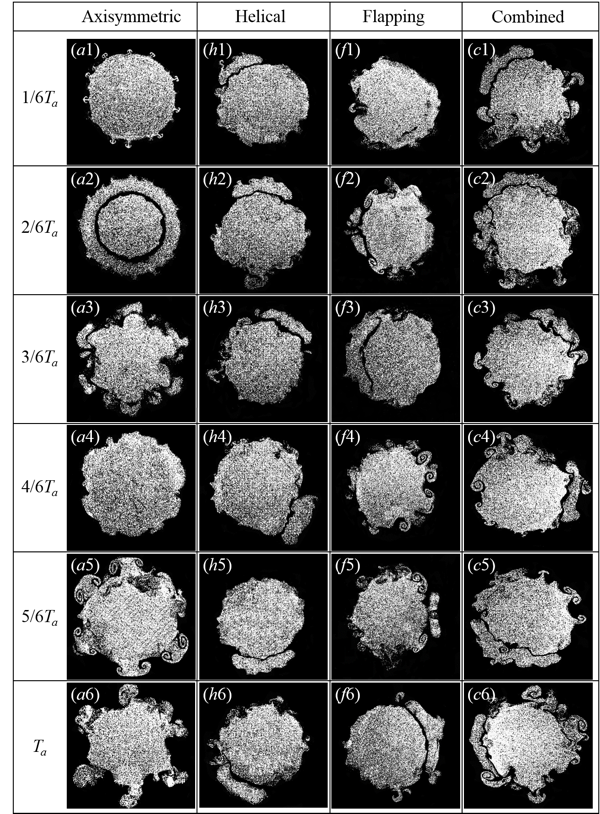}\\
    \caption{Sequential photographs of the cross-sectional flow structure captured at $x^\star$ = $0.25$. 
    From top to bottom: six instances at $t_i$ = $iT_a/6$ ($i=1, 2\ldots, 6$) in one actuation period $T_a$ (= $1/f_a$).
}
    \label{Fig:Crossplane}
 \end{figure}
%---------------------------------------------------------------

%--- Figure ------------------------------------------------------------
\begin{figure}
    \centering
    \includegraphics[width=0.9\textwidth]{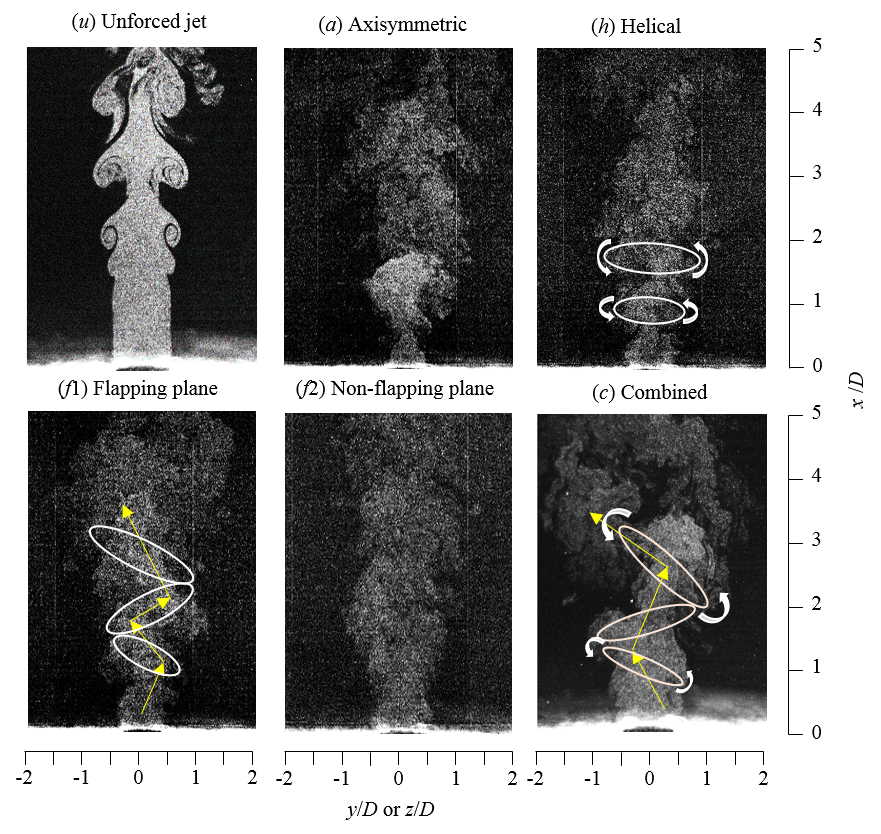}\\
    \caption{Flow visualization of 
    ($u$) unforced jet 
    ($a$) axisymmetric, 
    ($h$) helical, 
    ($f$) flapping, and
    ($c$) combined forcings 
    from the figure \ref{Fig:LearningCurve}, respectively. 
    White ellipses and crooked arrows indicate vortex rings and their rotation, respectively, 
    and yellow arrows highlight the jet column oscillation.
}
    \label{Fig:Streamwiseplane}
 \end{figure}
%---------------------------------------------------------------

The control mechanisms may be elucidated from the analysis of the spatio-temporal actuations, shown in figure \ref{Fig:Actuation}, 
extracted from the control laws of $n$ = 1, 2, 5 and 11. 
Each circular pie corresponds to one sixth of the excitation period, 
while its six sectors represent the six minijets. 
The arrow indicates that the minijet is on and the radial depth of 
the blue area is proportional to the duration 
when the minijet is injecting. 
The spatio-temporal actuation is found to be internally consistent with the cross-sectional and streamwise flow structure shown in figures \ref{Fig:Crossplane} and \ref{Fig:Streamwiseplane}. 
Axisymmetric forcing (figure \ref{Fig:Actuation}$a1-a6$) is characterized 
by simultaneous blowing of all minijets and a small $\alpha$ of 13.3\%. 
As a result, the cross-sectional flow structure (figure \ref{Fig:Crossplane}$a1-a6$) is axisymmetric, 
and the ring vortex is evident. 
\citet{Yang2016jfm} discussed in detail the distortion, 
formation of longitudinal structures and generation of mushroom-like structures 
in the braid region between ring vortices, 
and presented a scenario on how the interactions between the longitudinal structures, 
mushroom-like structures and ring vortices enhance entrainment and mixing. 
The six synchronized minijet excitations greatly strengthen 
the ring vortices, as shown in figure \ref{Fig:Streamwiseplane}$a$ (cf. figure \ref{Fig:Streamwiseplane}$u$). 
For helical forcing (figure \ref{Fig:Actuation}$h1-h6$), 
two or three minijets are blowing simultaneously at any instant, 
with $\alpha$ reaching 40\%. 
These blowing actions rotate clockwise from ($h1$) to ($h6$). 
The greatly increased $\alpha$, 
probably required for the generation of helical motion, 
may act to inhibit the occurrence of mushroom-like structures 
\citep{ARUN2018jfm}, 
which are absent in figure \ref{Fig:Crossplane}($h1-h6$). 
The jet (figure \ref{Fig:Streamwiseplane}$h$) 
exhibits more spread than axisymmetric forcing 
(figure \ref{Fig:Streamwiseplane}$a$). 
In case of flapping forcing (figure \ref{Fig:Actuation}$f1-f6$), 
three adjacent minijets are blowing simultaneously 
at one instant with $\alpha$ = 13.3\% and 
are switched to the other three after a phase shift of $\pi$, 
thus creating the asymmetric flapping jet column 
(figures \ref{Fig:Crossplane}$f1-f6$, \ref{Fig:Streamwiseplane}$f1-f2$). 
The actuation configuration of the rightmost column ($n$ = 11) 
in figure \ref{Fig:Actuation} is complex. 
Firstly, the number of the simultaneously injecting minijets 
can be 1, 2, 3 or 4, mostly adjacent to each other. 
Secondly, the injecting minijets tend to rotate clockwise. 
The resulting effect is to produce both helical and flapping motions 
(figures \ref{Fig:Crossplane}$c1-c6$, \ref{Fig:Streamwiseplane}$c$). 
For convenience, hereinafter we refer to this forcing as the combined mode. 
Thirdly, $\alpha$ varies from one blowing minijet to another, 
from about 6.7\% to 53\%. 
The effect could be twofold. 
On one hand, the varying duty cycles of the blowing minijets 
may yield a resultant blowing force not going through 
the jet centreline (figure \ref{Fig:Actuation}$c1-c6$). 
This may produce a precession effect, 
causing additional jet column oscillation \citep{Wong2004eif}. 
On the other hand, 
whilst a small $\alpha$ facilitates the generation of 
mushroom-like structures (figure \ref{Fig:Crossplane}$c1-c6$), 
a large value enhances the strength of the flapping motion 
(figure \ref{Fig:Streamwiseplane}$c$), as noted by \citet{ARUN2018jfm}. 
All the features, confirmed by more detailed analysis in section 6, act to promote mixing, consistent with the observed minimum $J$ (figure \ref{Fig:LearningCurve}). 
The generation of such a sophisticated control mode, 
along with the generation of a complex turbulent flow structure, 
would have been extremely challenging for conventional control techniques, 
be it open- or closed-loop and model-based or model-free approaches.

\subsection{Control landscape: cartographing all actuation laws}
\label{ToC:Control landscape: cartographing all actuation laws}

%--- Figure ------------------------------------------------------------
\begin{figure}
    \centering
    \includegraphics[width=0.9\textwidth]{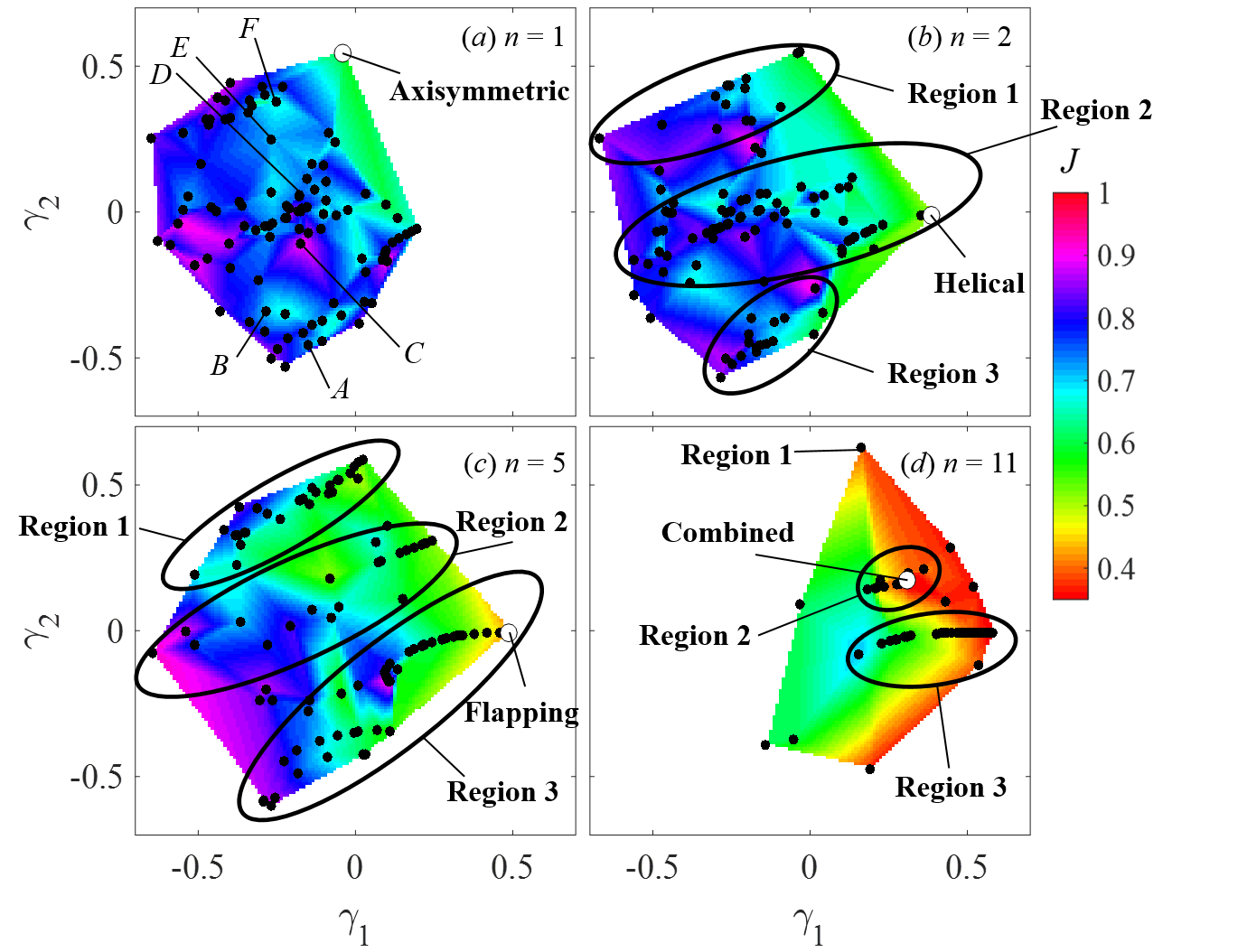}\\
    \caption{Control landscape associated with generations 1, 2, 5 and 11 (400 individuals). 
    Each symbol represents an individual control law. 
    The color scheme corresponds to the cost value $J$ of the control laws, e.g. 
    ($a$) $n$ = 1, ($b$) $n$ = 2, ($c$) $n$ = 5, ($d$) $n$ = 11. 
    The white circle of bigger size corresponds to 
    the best individual of one generation. 
    The elliptic contours enclose similar control laws.
    }
    \label{Fig:ProximityMap}
\end{figure}
%-----------------------------------------------------------------------

Proximity maps (\S~\ref{Toc:Parameters and control landscape}) provide 
a very revealing illustration of the evolution process of the control laws 
The underlying metric between two control laws $\boldsymbol{b}$ and $\boldsymbol{b'}$ is given by $D_{ij}$ (Eq. \ref{Eqn:Distance}). 
Figure \ref{Fig:ProximityMap} presents the proximity map of the control laws 
in a two-dimensional plane such that this metric is optimally preserved. 
This plane is spanned by the feature coordinates $\gamma_1$ and $\gamma_2$, 
which are derived from a mathematical optimization process. 
The details of deriving $\gamma_1$ and $\gamma_2$from individual generations are given in \citet{Cox2001MS}. 
Physically, the distance between two points, 
which are given in terms of ($\gamma_1$, $\gamma_2$), 
in the plane is directly linked to the extent 
how closely similar to each other two control laws are. 

The subfigures display the feature coordinates of the four discussed generations ($n$ = 1, 2, 5 and 11). 
Interesting observations and inferences could be made from the subfigures. 
For $n$ = 1 (figure \ref{Fig:ProximityMap}$a$), 
the points appear rather randomly distributed. 
The optimal solution of axisymmetric forcing, 
indicated by an open circle, 
occurs at ($\gamma_1$, $\gamma_2$) = (- 0.04, 0.54), 
where $\gamma_2$ is the largest of all points. At n = 2 (figure \ref{Fig:ProximityMap}$b$), 
these points appear forming three separated regions numbered 1 through 3, 
as enclosed by the elliptic contours, 
where most of the control laws fall in. 
The optimal solution that appears in Region 2 
and corresponds to helical forcing now occurs at 
($\gamma_1$, $\gamma_2$) = (0.39, -0.01), 
where $\gamma_1$ is the largest of all points. 
At $n$ = 5, the individual points tend to populate along discrete curves 
(figure \ref{Fig:ProximityMap}$c$), 
which is a commonly observed phenomenon of the AI control, 
e.g. \citet{Li2017aom} and \citet{Wu2018eif}. 
The optimal solution now takes place at ($\gamma_1$, $\gamma_2$) = (0.49, -0.01), 
where $\gamma_1$ is again the largest of all, 
within Region 3 and corresponds to flapping forcing. 
It has been confirmed that most of the data points within 
Regions 1, 2 and 3 correspond to axisymmetric, helical, and flapping forcings, respectively. 
By $n$ = 11 (figure \ref{Fig:ProximityMap}$d$), 
the optimal solution occurs at ($\gamma_1$, $\gamma_2$) = (0.31, 0.17) within Region 2, 
corresponding to the combined mode. 
It is noteworthy that those points in this region now correspond to either helical or combined forcing; 
this very fact may suggest that the so-called combined forcing 
may have been developed from and more closely related to helical forcing. 
Interestingly, Region 1 where axisymmetric forcing takes place 
contracts to a single point, while Region 2 or 3 is reduced to a curve.

The cost $J$ associated with each point is color-coded from red 
($J$ = 0) to pink ($J$ = 1) as indicated by the color bar. 
The cost values within each colored area are interpolated from the 100 individuals in each generation. 
The control landscape at $n$ = 1 displays quite a few minima, 
as indicated by green colored areas, in the feature plane (figure \ref{Fig:ProximityMap}$a$), 
suggesting the complexity of the learning task in the early stage, 
internally consistent with the rather random distribution of the points. 
The landscape becomes simpler towards 
the rightmost boundary of the generation. 
With increasing number of generations, 
the individuals tend to line up on the 
ridge-curves marking the cost valleys.

%--- Figure ------------------------------------------------------------
\begin{figure}
    \centering
    \includegraphics[width=0.9\textwidth]{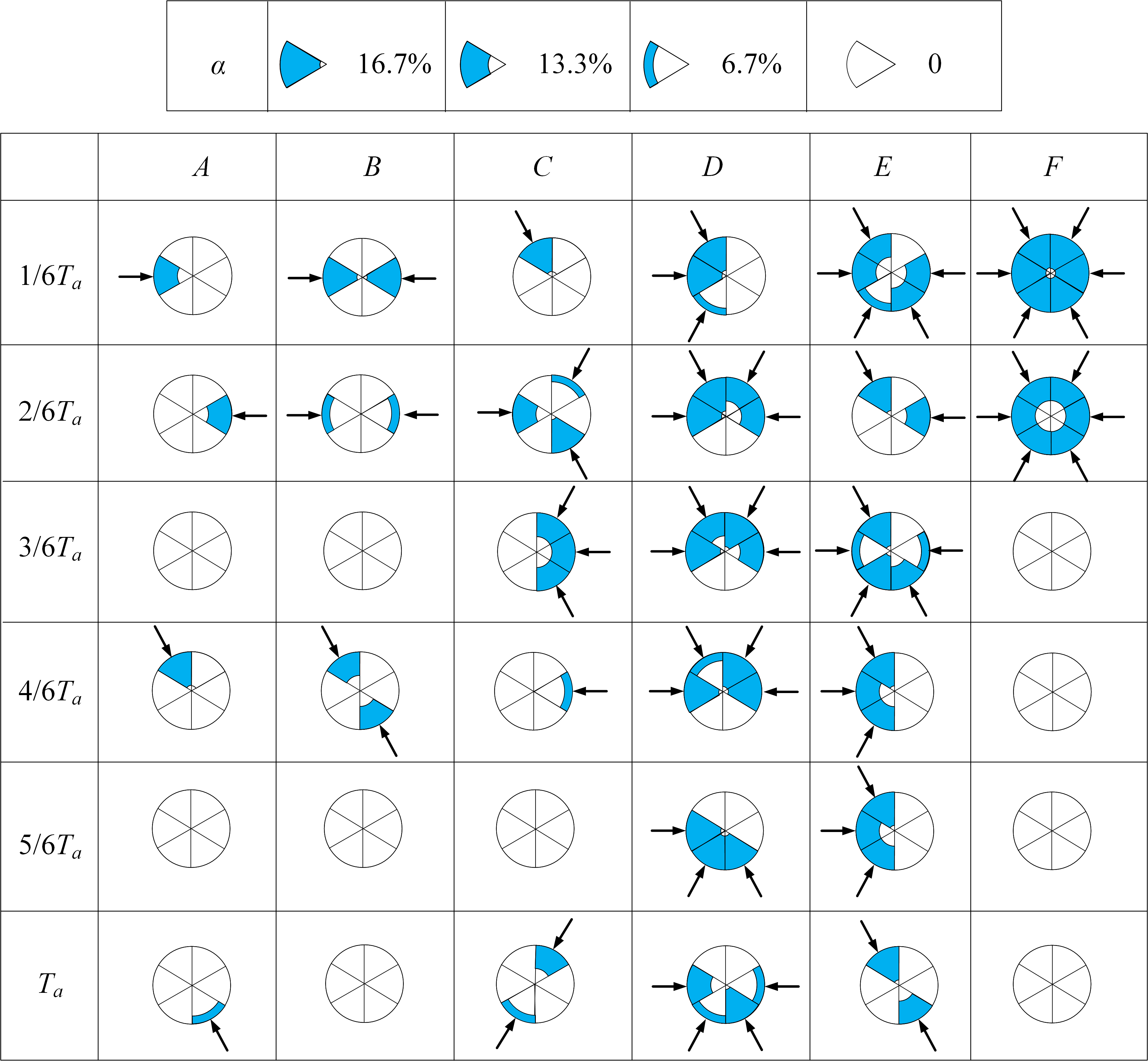}\\
    \caption{Actuation associated with control laws $A-F$, 
    extracted from figure \ref{Fig:ProximityMap}$a$. 
    The symbols are as in figure \ref{Fig:Actuation}.
    }
    \label{Fig:AI actuation}
\end{figure}
%----------------------------------------------------------------------

The feature coordinates have technically no a priori meaning. 
However, an approximate physical meaning of $\gamma_1$ and $\gamma_2$  may be inferred 
from the careful analysis of the control laws. 
The coordinate $\gamma_1$ is correlated with the degree of asymmetric forcing, 
while $\gamma_2$ is linked to the number of simultaneously injecting minijets. 
Consider six arbitrarily chosen control laws $A-F$ in the first generation (figure \ref{Fig:ProximityMap}$a$). 
Figure \ref{Fig:AI actuation} shows the on- and off-states of the minijets 
corresponding to the six control laws. 
Evidently, the maximum number of simultaneously injecting minijets 
increases from one (law $A$) to six (law $F$).

\section{Discussion: representative control laws and flow structures}
\label{ToC:Flow aspect}

\subsection{Jet spread and predominant flow structures}
\label{ToC:Jet spread and predominant flow structures}

%--- Figure ------------------------------------------------------------
\begin{figure}
 	\centering
    \includegraphics[width=0.9\textwidth]{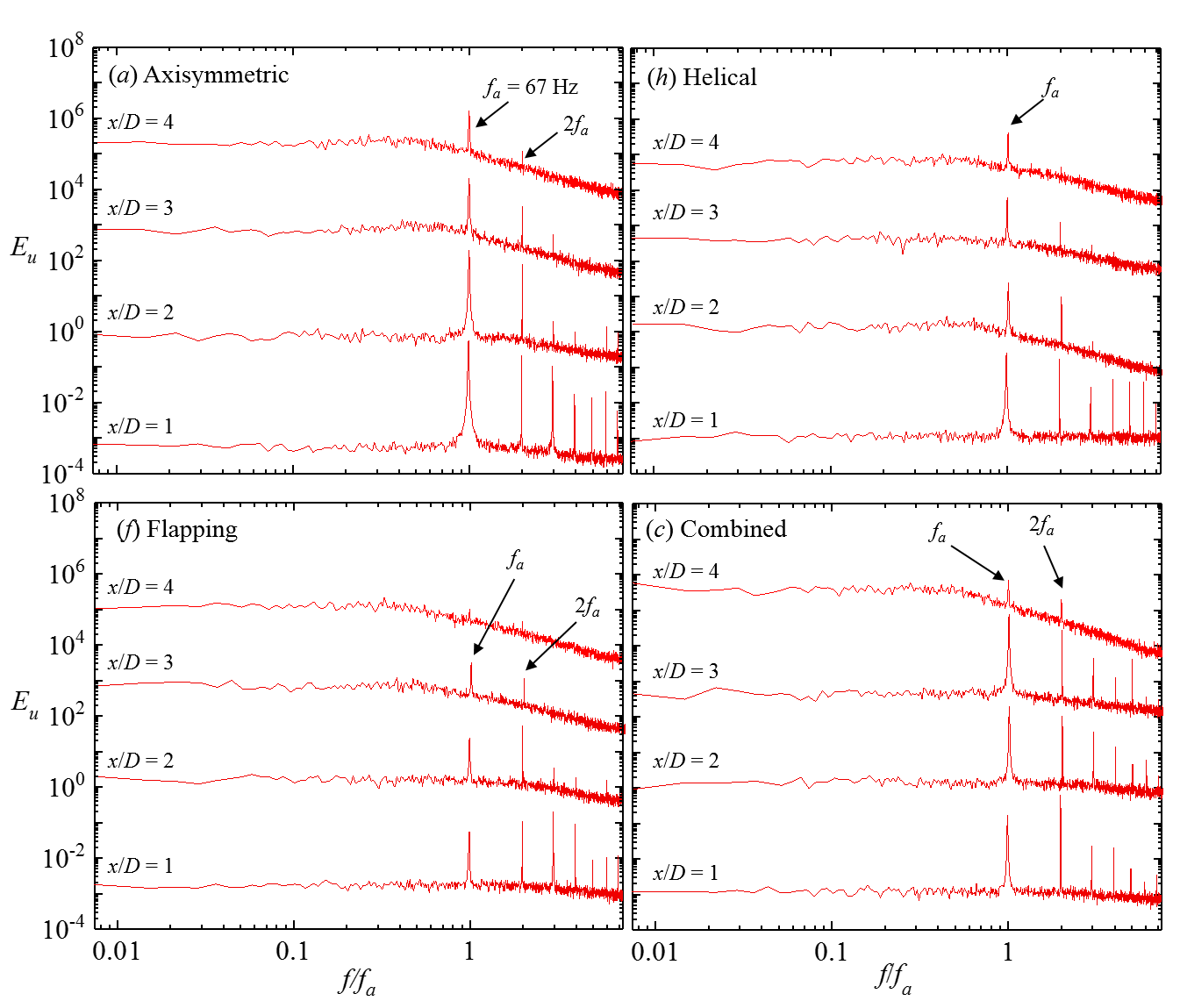}\\
    \caption{Normalized streamwise power spectra of fluctuating velocity 
    $u$ measured on the centerline in the presence of main jet: 
    ($a$) axisymmetric, 
    ($h$) helical, 
    ($f$) flapping and 
    ($c$) combined forcings.}
 	\label{Fig:PSD}
 \end{figure}
%----------------------------------------------------------------------

\begin{figure}
    \centering
    \includegraphics[width=1\textwidth]{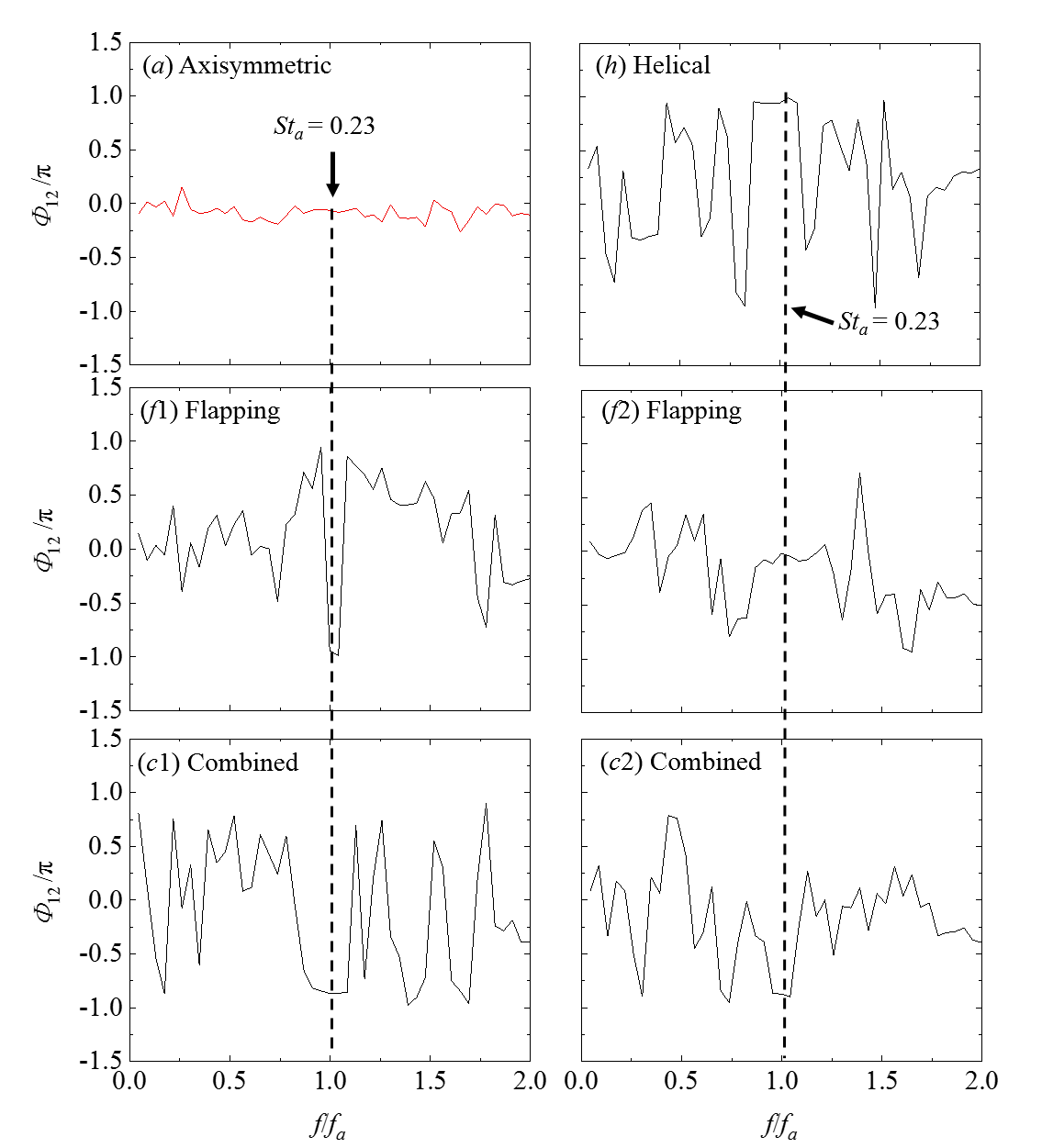}\\
    \caption{ Phase spectrum $\Phi_{12}$ between two streamwise fluctuating velocity signals $u$ from hot-wires measured at $x/D = 1$ and $y/D$ or $z/D$ = $\pm 0.3$. 
    $S_{ta}$ = $f_aD/\overline U_j$.
    }
    \label{Fig:Phase_spectrum}
 \end{figure}

Axisymmetric forcing at $n$ = 1 (figure \ref{Fig:Streamwiseplane}$a$) 
leads to an early disruption of the potential core 
and a significantly improved entrainment, 
as suggested by the substantially reduced `neck' due to ambient fluid 
(dark color) brought into the jet by the greatly increased strength 
of vortices near the nozzle exit. 
The flow structure exhibits an appreciably increased lateral spread than the unforced jet (figure \ref{Fig:Streamwiseplane}$u$). 
The $E_u$ of the centreline $u$ (figure \ref{Fig:PSD}$a$) 
measured at $x/D$ = 1.0 displays the most pronounced peaks of all 
at $f/f_a$ = 1.0 and its harmonics, 
though the peaks at higher harmonics vanish rapidly 
from $x/D$ = 1.0 to 4.0. 
These peaks become very sharp because of the minijet excitation. 
The ring vortices remain axisymmetric (figure \ref{Fig:Crossplane}$a1-a6$), 
which is corroborated by the spectral phase $\Phi_{12}$, 
about zero over the entire range of $f/f_a$ = $0 - 2.0$, 
between two simultaneously measured hotwire signals $u_1$ and $u_2$ 
at $x/D$ = 1 and $y/D$ = $\pm$0.3 (figure \ref{Fig:Phase_spectrum}$a$). 
The $\Phi_{12}$ is calculated by $tan^{-1}(Q_{12}/Co_{12})$, 
where $Co_{12}$ and $Q_{12}$ represent the cospectrum and quadrature spectrum of $u_1$ and $u_2$, respectively \citep{zhou2002jfm}. 

An even larger lateral spread is achieved for helical forcing 
at $n$ = 2 (figure \ref{Fig:Streamwiseplane}$h$) 
due to the rotating flow structure (figure \ref{Fig:Crossplane}$h1-h6$). 
The $E_u$ measured on the centreline shows less pronounced peaks 
at $f/f_a$ = 1.0 and its harmonics than its counterpart 
for axisymmetric forcing. 
This is because there are only two or three injecting minijets 
at any instant for helical forcing (figure \ref{Fig:Actuation}$h1-h6$), 
which produce considerably less velocity fluctuations at the jet centre 
than six minijets (figures \ref{Fig:exit_PSD} and \ref{Fig:exit_urms}). 
Further, the peaks at $f/f_a \ge 2.0$ disappear at $x/D \ge 1$. 
The $\Phi_{12}$ is about $\pi$ over a range of frequencies about 
$f/f_a$ = 1.0 (figure \ref{Fig:Phase_spectrum}$h$), 
as is expected based on figure \ref{Fig:Actuation}($h1-h6$). 

%--- Figure ------------------------------------------------------------
%-----------------------------------------------------------------------

At $n$ = 5, the jet column wobbles right and left, 
as indicated by yellow arrows in figure \ref{Fig:Streamwiseplane}$f1$, 
in the flapping plane (the $x-y$ plane) but not 
in the orthogonal $x-z$ plane (figure \ref{Fig:Streamwiseplane}$f2$), 
which is symmetrical about $y$ = 0. 
The lateral spread appears exceeding appreciably that for helical forcing, 
echoing the considerably improved mixing shown in figure \ref{Fig:LearningCurve}. 
Interestingly, the peaks at $f/f_a$ = 1.0 and its harmonics in $E_u$ 
are less pronounced than their counterparts of helical forcing 
(figure \ref{Fig:PSD}). 
The peaks of the former decay more rapidly, 
completely vanished by $x/D$ = 4.0, than the latter. 
The observation is consistent with the perception that it is the flapping motion, 
not the large-scale vortices, that plays a predominant role 
in enhancing mixing in this case. 
The flapping motion of the jet is characterized by a negative correlation 
between the two fluctuating streamwise velocities obtained 
on the opposite side of the jet \citep{Goldschmidt1973pof}. 
Indeed, $\Phi_{12}$ is about $\pi$ over a very narrow frequency band 
about $f/f_a$ = 1.0 in the flapping plane but zero over 
a rather broad range of frequencies in the non-flapping plane 
(figure \ref{Fig:Phase_spectrum}$f1, f2$), as observed by \citet{Yang2016jfm}. 
Note that the peak at $f/f_a$ = 2.0 is larger than at $f/f_a$ = 1.0 
for flapping forcing (figure \ref{Fig:PSD}$f$). 
This behaviour is ascribed to the flapping motion (figure \ref{Fig:Streamwiseplane}$f1$) caused by 
two separate excitations with a phase shift of $\pi$ 
within each excitation cycle (figure \ref{Fig:Actuation}), 
which are captured by the hotwire. 

The combined mode is distinct from all other forcings. 
Firstly, its spread shown in figure \ref{Fig:Streamwiseplane}($c$) 
is clearly the largest of all, due to the presence of both flapping 
and helical motions (figures \ref{Fig:Actuation}$c1-c6$ and \ref{Fig:Crossplane}$c1-c6$), 
internally consistent with the smallest $J$ in figure \ref{Fig:LearningCurve}. 
Secondly, its $E_u$ (figure \ref{Fig:PSD}$c$) displays 
a number of differences from other forcings. 
The peak at $f/f_a$ = 1.0 grows in amplitude from $x/D$ = 1 to 3, 
while its counterpart for other three forcings all decay quickly. 
Furthermore, the peaks at the higher harmonics of $f/f_a$ = 1.0 
decay little for the same range of $x/D$, in distinct contrast to 
their counterparts of other forcings where these peaks retreat rapidly. 
Naturally, compared with other three forcing modes, 
there are many peaks at the higher harmonics of $f/f_a$ = 1.0, especially at $x/D$ = 3. 
When manipulating main jet using a single unsteady minijet, 
\citet{ARUN2018jfm} made a similar observation, 
which was ascribed to the use of a small duty cycle. 
The small duty cycle occurs in every phase of the combined mode 
(figure \ref{Fig:Actuation}$c1-c6$). 
Another note is that the peak at $f/f_a$ = 2.0 is larger than 
at $f/f_a$ = 1.0 for $x/D$ = 1 and remains very pronounced downstream, 
similarly to the flapping forcing case (figure \ref{Fig:PSD}$f$). 
Thirdly, its $\Phi_{12}$ (figure \ref{Fig:Phase_spectrum}$c$) approaches 
anti-phase at $f/f_a$ = 1.0 for both orthogonal planes examined. 
A rather broad plateau occurs about $f/f_a$ = 1.0, 
where $\Phi_{12}$ $\approx$ -0.86 $\pi$. 
This phase shift differs appreciably in value from flapping or helical forcing where $\Phi_{12}$ $\approx$ $\pi$, 
and is probably connected to the presence of the oscillating component 
in this mode. 
Note that the combined mode produces the nearly anti-phased behavior 
for all planes through the $x$ axis. 
However, in flapping forcing, this anti-phased behavior takes place 
only in the flapping plane.

\subsection{Momentums impinging upon main jet and jet centre trajectory}
\label{ToC:Momentums impinging upon main jet and jet centre trajectory}

%--- Figure ------------------------------------------------------------
\begin{figure}
    \centering
    \includegraphics[width=0.95\textwidth]{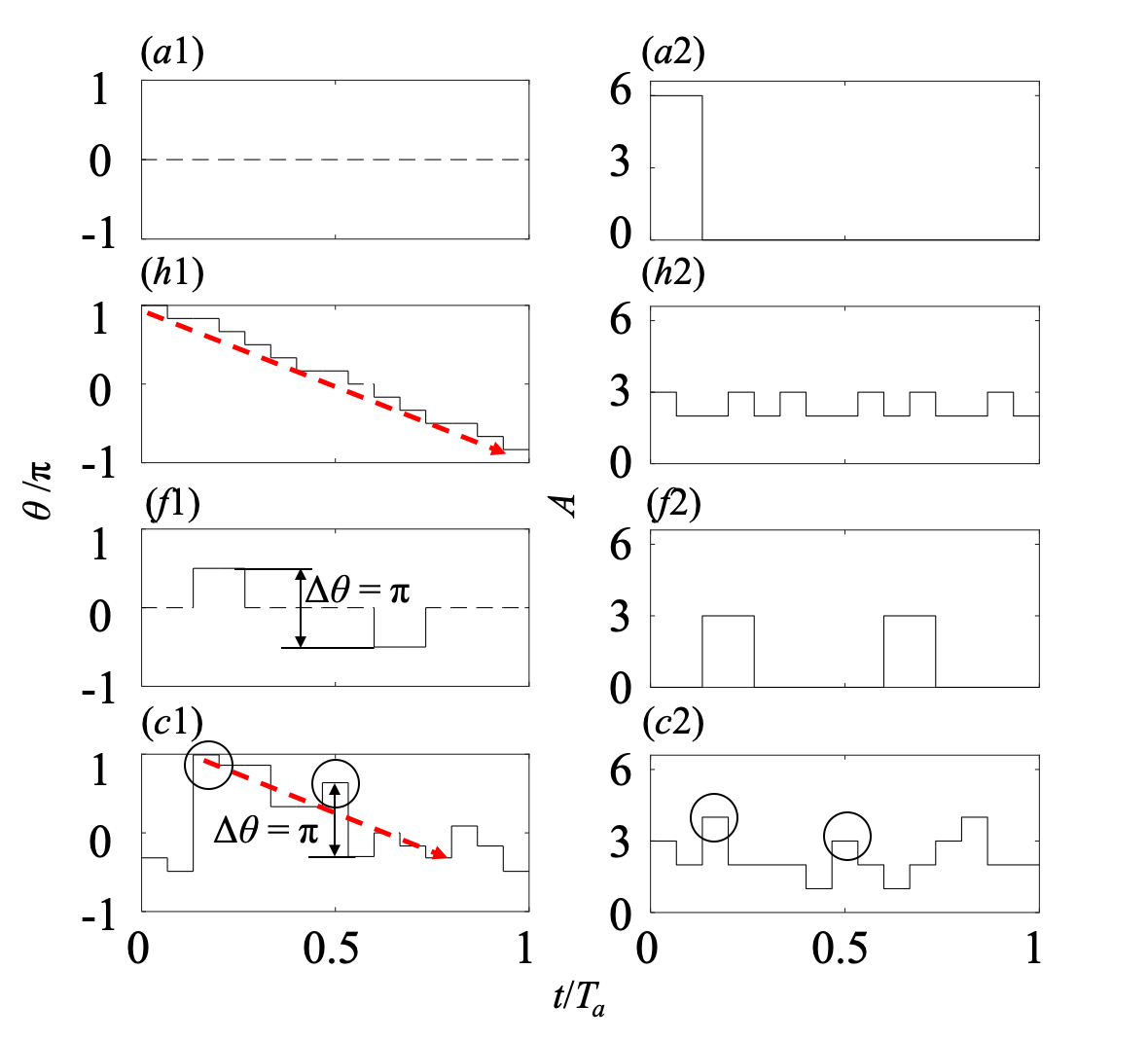}\\
    \caption{The angle $\theta$ (left column) and magnitude 
    $A$($t$) (right column) of 
    the effective minijet actuation vector $\boldsymbol{Q}(t)$ 
    for the best control law of $n$ = 1, 2, 5 and 11).}
    \label{Fig:angle}
 \end{figure}
%---------------------------------------------------------------

Additional insight may be gained into the flow physics of 
the combined mode by examining the sum of the momentums 
due to individual injecting minijets impinging upon the main jet 
and the direction of their resultant momentum. 
As the averaged mass flux is the same for all minijets, 
the maximum actuation velocity scales roughly with the inverse of 
the duty cycle $\alpha$, as demonstrated in figure \ref{Fig:1mj_signal}. 
As such, the maximum actuation velocity $U_{a,i}$ of the $i$th minijet 
is proportional to the product of $b_i(t)$ and $1/\alpha_i$, viz.

\begin{equation}
U_{a,i}(t) \propto \frac{b_i(t)}{\alpha_i},
\label{Eqn:U_a}
\end{equation}
where $b_i(t)$ is a signal generated by 
$K_i(\boldsymbol{s}(t), \boldsymbol{h}(t))$, 
given in eq. \ref{Eqn:ControlLaw}, 
its values 0 and 1 corresponding to the `on' and `off' states, respectively. 
The time-averaged mass flow rate is the same, 1.2\%, for every minijet.  
The sum of the momentums due to individual injecting minijets 
impinging upon the main jet is parameterized by

\begin{equation}
A(t)=\sum_{i=1}^{N} \frac{b_i(t)}{\alpha_i}.
\label{Eqn:amplitudeion}
\end{equation}
By definition, the time-averaged amplitude is the number 
of active minijets, i.e. 0 for unforced flow and $\overline{A(t)}$ = $N$ = 6 
for actuation, implying a total time-averaged mass-flow rate of 7.2\%. 
On the other hand, the resultant momentum vector of the momentums 
associated with individual injecting minijets is given by 

\begin{equation}
\boldsymbol{Q}(t)=\sum_{i=1}^{N} \frac{b_i(t)}{\alpha_i} \boldsymbol{e}_i,
\label{Eqn:amplitude}
\end{equation}
where $\boldsymbol{e}_i$ = -($\cos \theta_i$, $\sin \theta_i$) is 
a unit vector in the direction of the $i$th minijet located 
at angle $\theta_i$ in the $y-z$ plane. 
The minus sign refers to an inward velocity towards the jet centre. 
$N$ is the number of injecting minijets.

Figure \ref{Fig:angle} shows the variation in angle $\theta(t)$ of $\boldsymbol{Q}(t)$, 
with respect to the $y$ axis, and $A(t)$ with time $t$ over 
two actuation periods $T_a$ for the axisymmetric, helical, flapping 
and combined forcing modes ($n$ = 1, 2, 5 and 11). 
For axisymmetric forcing, $\theta(t)$ is undetermined as $\boldsymbol{Q}(t)$ vanishes identically, 
as $b_1$ = $b_2$ =$\ldots$= $b_6$ and $\sum \boldsymbol{e}_i$ = 0 (
figure \ref{Fig:Actuation}$a1$). 
Therefore, $A(t)$, albeit large (figure \ref{Fig:angle}$a2$), 
would not make the jet column oscillate (figure \ref{Fig:Streamwiseplane}$a$). 
In the case of helical forcing, $\theta$ varies essentially linearly 
with $t$, as indicated by the red dashed line (figure \ref{Fig:angle}$h1$). 
The stepwise behaviour is caused by the discontinuous on-off actuation 
$b_i(t)$. 
A large $A(t)$ or one half of the strength of axisymmetric forcing 
(figure \ref{Fig:angle}$a2, f2$) occurs at $t/T_a$ = 0.2 and 0.7 
for flapping forcing, which correspond to a phase shift of $\pi$ 
(figure \ref{Fig:angle}$f1$). 
The behaviours of both $A(t)$ and $\theta(t)$ are fully consistent 
with our understanding of axisymmetric, helical and flapping forcings, 
thus providing a validation for applying $A(t)$ and $\theta(t)$ to describe the forcing on main jet.

For the combined mode, the variations in both $A(t)$ and $\theta(t)$ 
with $t/T_a$ are more complicated. 
Nevertheless, a number of features can be identified. 
Firstly, after reaching the first maximum as highlighted by a circle, 
$\theta(t)$ decreases, albeit not monotonically, over a duration of 1.3$\pi$ (figure \ref{Fig:angle}$c1$), similarly to helical forcing, 
as indicated by the red arrow. 
This feature implies a swirling forcing on main jet. 
Secondly, the maxima of A exceed those of helical forcing, 
suggesting a stronger swirl actuation. 
Thirdly, the phase shift between the second local maximum $\theta(t)$ 
and a local minimum $\theta(t)$ at $t/T_a$ = 0.67 is $\pi$, 
as highlighted by the vertical arrow in figure \ref{Fig:angle}$c1$, 
pointing to the signature of flapping forcing. 
The local maximum $A$ at $t/T_a$ = 0.67 (figure \ref{Fig:angle}$c2$) 
is associated with a phase change of $\pi$ (figure \ref{Fig:angle}$c1$), 
highlighted by a circle. 
All the observations suggest that the combined mode is rather unique, 
featured by both helical and flapping motions, and further 
by the stronger strengths of flapping and vortical motions than other cases, 
thus accounting for the largest entrainment and mixing of all. 

%--- Figure ------------------------------------------------------------
\begin{figure}
    \centering
    \includegraphics[width=0.9\textwidth]{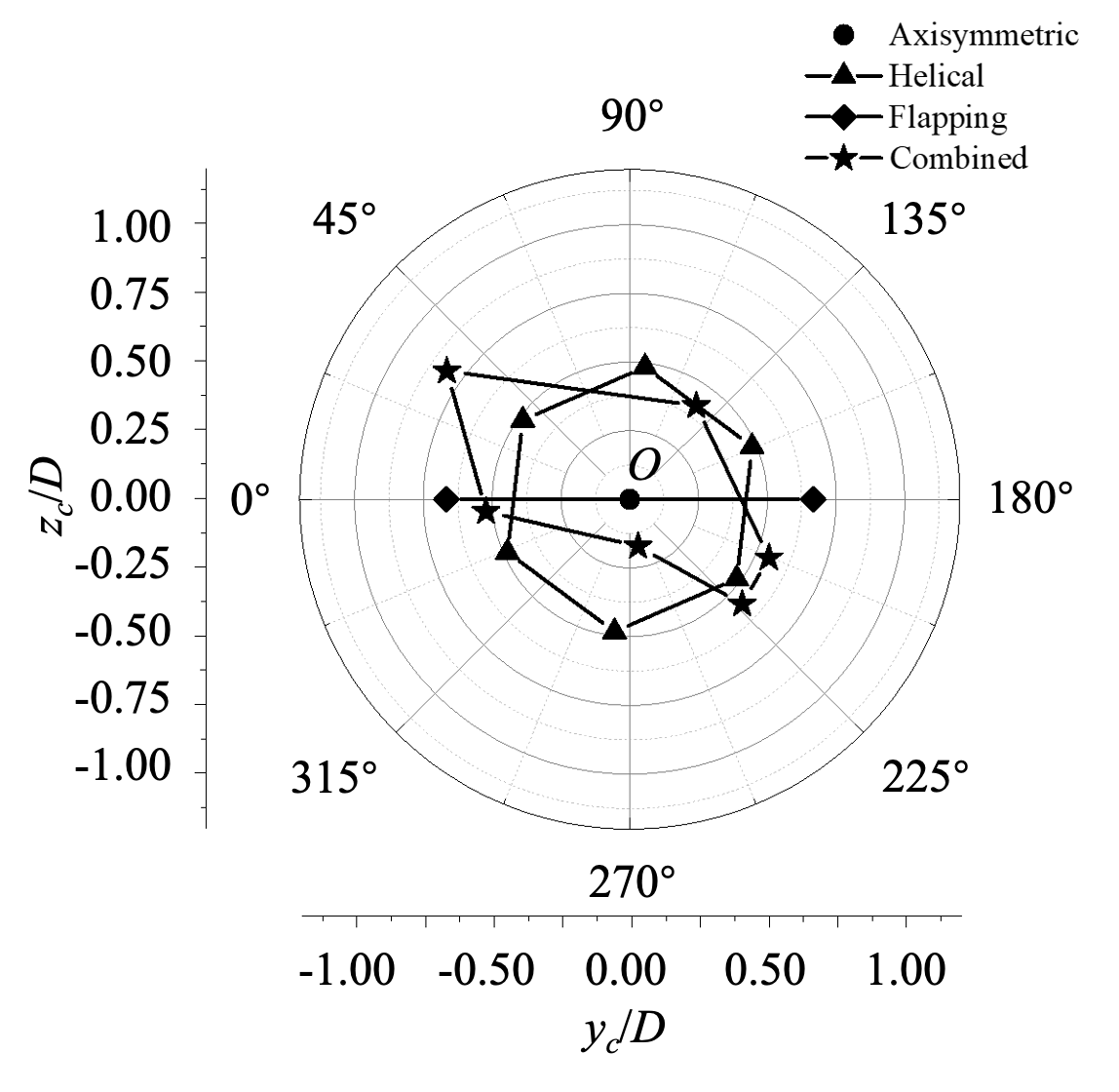}\\
    \caption{Jet centre trajectory ($y_c, z_c$) predicted 
    from the resultant momentum vector $\boldsymbol{Q}(t_i)$ of 
    the momentums due to individual injecting minijets 
    at phases $t_i$ = $iT_a/6$ ($i = 1, 2\ldots, 6$).}
    \label{Fig:Jet centre}
 \end{figure}
%---------------------------------------------------------------

Evidently, the jet centre $\boldsymbol{r}_c$ = ($y_c, z_c$) responds to 
$\boldsymbol{Q}(t_i)$ (eq. \ref{Eqn:amplitude}), 
and it is plausible to assume the jet centre trajectory to be correlated 
with $\boldsymbol{Q}(t_i)$. 
This centre can be characterized as the `centre of gravity' of the streamwise velocity in a cross-stream plane of $x$ = constant:

\begin{equation}
\boldsymbol{r}_c(t)=\iint d\boldsymbol{r} \, \boldsymbol{r} \, u(\boldsymbol{r},t)/\iint d\boldsymbol{r} \, u(\boldsymbol{r},t).
\label{Eqn:jet centre}
\end{equation}
Here, $\boldsymbol{r}$ = ($y, z$) represents the coordinate 
in the cross-stream plane. 
For simplicity, the $x$-dependency of ($y_c, z_c$) 
will be dropped out hereafter. 
Over one excitation period with a time step $\triangle t$ = $T$/6, 
the jet centre takes six positions, i.e. 
$\boldsymbol{r}_c(t_i)$ = ($y_c(t_i), z_c(t_i)$), $t_i$ = $i \triangle t$, 
$i = 0, 1\ldots, 5$. 
We make the most simple assumption that the jet centre displacement 
($\triangle y_c, \triangle z_c$)($t_i$) = 
($y_c(t_{i+1}) - y_c (t_i)$, $z_c(t_{i+1}) - z_c(t_i)$) 
over $\triangle t$ is proportional to the actuation momentum, viz.

\begin{equation}
\boldsymbol{r}_c(t_{i+1})=\boldsymbol{r}_c(t_i)+\boldsymbol{Q}(t_i),
\label{Eqn:centre add}
\end{equation}
for i = $0,\ldots, 5$, 
where the proportionality constant is set to unity again for simplicity. 
The discrete time dynamics \ref{Eqn:centre add} can be considered to 
a rough discretization of the dynamics equation 
$d(\boldsymbol{r}_c)/dt$ = $\boldsymbol{Q}$, 
which describes the jet centre motion under the external momentum. 
Eq. \ref{Eqn:centre add} represents 10 equations for 12 unknown jet centre coordinates. 
The remaining equations are obtained from the observation that 
the control law is periodic in time and the time-averaged 
actuation momentum vanishes based on experimental constraints, 
i.e. the same averaged mass flow through each minijet.  
Hence, the average jet centre position can be expected to vanish: 

\begin{equation}
\sum_{i=0}^5\boldsymbol{r}_c(t_{i})=0.
\label{Eqn:centre zero}
\end{equation}
Equations \ref{Eqn:centre add} and \ref{Eqn:centre zero} constitute 12 linear equations for 12 unknowns, 
describing the motion of the jet centre over one excitation period. 
The jet centre dynamics is most easily solved by starting the integration 
at the origin with $x_c(t_0)$ = $y_c(t_0)$ = 0, 
iteratively computing the positions at $t_i$, $i = 1\ldots, 5$ 
with (eq. \ref{Eqn:centre add}) and adding a translation consistent 
with vanishing averaged jet centre (eq. \ref{Eqn:centre zero}). 

Figure \ref{Fig:Jet centre} presents the trajectories of the jet centre 
within each excitation period for the four forcings, 
which are calculated based on the control laws shown in figure \ref{Fig:Actuation} or Eqs. (\ref{Eqn:generation_1}-\ref{Eqn:generation_11}) and (\ref{Eqn:centre add}-\ref{Eqn:centre zero}). 
Apparently, the jet centre vanishes identically for axisymmetric forcing 
where $\boldsymbol{Q}(t_i)$ $\equiv$ 0 and oscillates along 
the $y-$direction between two extremes in the $x–y$ plane 
for flapping forcing where $\boldsymbol{Q}(t_i)$ changes 
from the positive $y-$direction at one phase to the negative 
at next phase or vice versa. 
Helical forcing, i.e. a uniformly rotating $\boldsymbol{Q}$ vector, 
leads to a uniformly processing jet and the jet centre moves 
along a circle around the axis of symmetry. 
The result conforms to previous reports. 
\citet{koenig2016jfm} experimentally investigated the turbulent jet 
under the helical mode excitation and observed a precessing jet column 
when the helical structures were spatially amplified in the shear layer. 
\citet{ZHANG2016etfs} found in a similar experiment that the jet centreline 
under helical excitation was offset slightly and precessed 
around the initial axis of the core flow. 
For the combined forcing, the motion of the jet centre is more complicated. 
Its trajectory is apparently ellipse-like, 
suggesting the occurrence of a precession jet. 
In contrast to helical forcing, the distance of the jet centre 
from the centre of symmetry varies, 
along with the separation between the centres of two consecutive phases, 
suggesting the speed of swirling changes with time. 
Furthermore, this ellipse-like path indicates an oscillating jet column, 
a feature of the flapping motion.

\subsection{Velocity field}
\label{ToC:Velocity field}

%--- Figure ------------------------------------------------------------
\begin{figure}
 	\centering
    \includegraphics[width=0.8\textwidth]{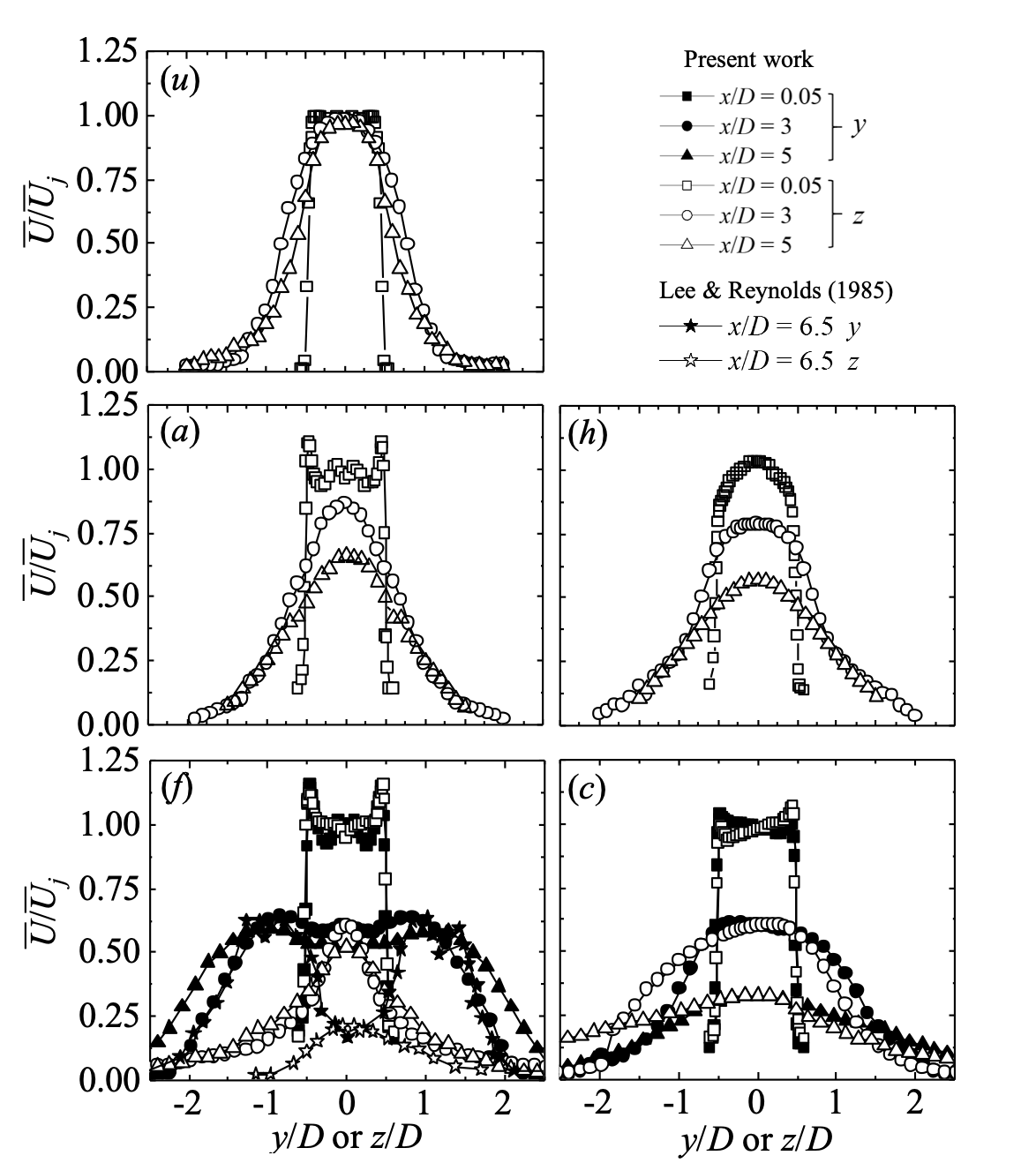}\\
    \caption{Radial distributions of $\overline U/\overline U_j$ 
    measured at different $x/D$ in both $x-z$ and $x-y$ planes for 
    ($u$) unforced jet; 
    ($a$) axisymmetric; 
    ($h$) helical; 
    ($f$) flapping, the bifurcation jet \citet{Lee1985} 
    is included for comparison; 
    ($c$) combimed.}
    \label{Fig:exit_U_profile}
 \end{figure}
%---------------------------------------------------------------

The velocity field may provide us with the crucial information on the flow structure. 
Figure \ref{Fig:exit_U_profile} presents the radial profiles of the hot-wire measured $\overline U/ \overline U_j$ at $x/D$ = 0.05. 
A number of observations could be made. 
Firstly, for all controlled cases except helical forcing, 
the mean velocity profile displays an overshoot at the nozzle exit, 
as noted by \citet{Wu2018eif} who used a single unsteady minijet 
to manipulate main jet. 
\citet{andreopoulos1984jfm} made the same experimental observation 
in case of a circular jet in a cross-stream. 
They explained that the cross-stream fluid acted like a partial cover over 
the jet exit, causing the jet flow to bend around and to accelerate 
so that the velocity of the bent-over jet was somewhat higher 
than the cross-stream velocity. 
Secondly, the widths of the mean velocity profiles under control 
are larger than that of the unforced jet, 
indicating that the shear layer grows laterally, 
and the main jet becomes wider right at the nozzle exit. 
Thirdly, the mean velocity profiles of the axisymmetric and helical forcing 
at jet exit are almost symmetrical about $y/D$ or $z/D$ = 0, 
in general with its maximum at the centre (figure \ref{Fig:exit_U_profile}$a, h$). 
The velocity profile of flapping forcing is also symmetric about $y/D$ and $z/D$ = 0 (figure \ref{Fig:exit_U_profile}$f$), 
though displaying two peaks off the centre in the $x-y$ or flapping plane 
and indicating the occurrence of bifurcation. 
As shown in figure \ref{Fig:Streamwiseplane}($f1$), two consecutive rings are locally connected, 
forming a zigzag flow structure, as observed in
\citet{Carlos2002pof} numerical investigation of a bifurcating jet 
(their figure 4$a$). 
Furthermore, the cross-flow distributions $\overline U/\overline U_j$ at 
$x/D$ = $3-5$ (figure \ref{Fig:exit_U_profile}$f$) display two peaks 
in the bifurcation ($x-y$) plane, 
while those in the bisection ($x-z$) plane show only one peak 
on the centreline. Also, the jet grows slowly in width 
in the bisection plane. 
All the features are similar to \citet{Lee1985} data 
where loudspeakers were used to produce a bifurcating water jet at $Re_D$ = 4300. 
The present data deviate quantitatively from \citet{Lee1985} 
as a result of distinct actuation techniques, experimental setups 
and $Re_D$ between the two investigations. 
Finally, the velocity profile under combined forcing mode (figure \ref{Fig:exit_U_profile}$c$) is distinct from others and a little tilted 
at the nozzle exit. 
Further downstream, $\overline U/\overline U_j$ remains 
asymmetrical about the jet centre but, 
unlike the flapping mode (figure \ref{Fig:exit_U_profile}$f$), 
does not show the twin-peak distribution. 
\citet{WONG2003etfs} produced a precession jet by issuing a jet 
into a cylindrical chamber with a small axisymmetric inlet at one end 
and an exit lip at the other. 
The inlet flow separates at the abrupt 
inlet expansion and reattaches asymmetrically to the wall of the chamber. 
Asymmetry of the flow within the chamber causes the reattaching flow to precess around the inner wall of the chamber, 
resulting in a precessing exit flow. 
An asymmetric and rotating pressure field is thus established 
so that the entire flow-field, including the emerging jet, 
precesses \citep{nathan1998jfm}. 
The precessing jet proves to be highly effective in increasing 
the near-field spreading. Interestingly, 
the present distributions of $\overline U/\overline U_j$ exhibit 
a similarity to their counterparts of the precession jet 
(please refer to figure 10 in \citep{WONG2003etfs}) 
at the nozzle exit and downstream development. 

%--- Figure ------------------------------------------------------------
\begin{figure}
 	\centering
    \includegraphics[width=0.8\textwidth]{./Figures/S6_jet_exit_Urms_profile.png}\\
    \caption{Radial distributions of $u_{rms}/\overline U_j$ 
    measured at different $x/D$ in both $x-z$ and $x-y$ planes for 
    ($u$) unforced jet; 
    ($a$) axisymmetric; 
    ($h$) helical; 
    ($f$) flapping; 
    ($c$) combined. 
    }
 	\label{Fig:exit_Urms_profile}
 \end{figure}
%-----------------------------------------------------------------------

The radial distributions of $u_{rms}/\overline U_j$ (figure \ref{Fig:exit_Urms_profile}) are reasonably symmetrical about $y/D$ or $z/D$ = 0 
under control, except under combined forcing. 
The $u_{rms}$ rises greatly at $x/D$ = 0.05 throughout the main jet, 
as compared to the uncontrolled jet (figure \ref{Fig:exit_Urms_profile}$u$) 
where $u_{rms}/\overline U_j$ = 0.3\% at $y/D$ = 0 and 2.5\% 
at $y/D$ $\approx$ $\pm$0.43 due to the shear layer. 
The peak in the shear layer becomes very pronounced. 
The result suggests a turbulent jet at the nozzle exit, 
which is internally consistent with flow visualization data (figure \ref{Fig:Streamwiseplane}), 
and the shear layer instabilities are significantly amplified. 
The $u_{rms}/\overline U_j$ for axisymmetric forcing in the centre region 
is largest of all, due to the simultaneous injection of six minijets 
into the main jet, which causes a strong disturbance 
in the central region (figure \ref{Fig:exit_Urms_profile}$a$). 
The $u_{rms}/\overline U_j$ distribution displays twin peaks for helical forcing (figure \ref{Fig:exit_Urms_profile}$h$). 
This is reasonable as helical forcing may produce a hurricane-like helical motion 
with a centre that is more stable than the surrounding motion. 
The $u_{rms}/\overline U_j$ at $x/D$ = 0.05 under flapping forcing 
in the $x-y$ plane is larger than in the $x-z$ plane in the shear layer 
(figure \ref{Fig:exit_Urms_profile}$f$), 
as observed by \citet{Hussain1989jfm} and \citet{Zaman1996jfm}. 
The twin-peak behavior is evident along the $z$ axis 
due to the flapping motion, but not so along the y axis. 
The $u_{rms}/\overline U_j$ distributions under combined forcing are 
asymmetrical about the centre at the jet exit (figure \ref{Fig:exit_Urms_profile}$c$), 
and again resemble the procession jet \citep{Mi2005pof}. 
Unlike the case under flapping forcing, 
the $u_{rms}/\overline U_j$ profiles under combined forcing display 
marked twin peaks along the $y$ and $z$ axes, 
where the right peak is more pronounced than the left, 
which is probably linked to the helical motion. 
Furthermore, the $u_{rms}/\overline U_j$ peaks at $x/D$ = 3 
are substantially higher along the $y$ axis than along the $z$ axis, 
which remains discernible at $x/D$ = 5. 
All the features have been observed in the precession jet. 
\citet{Mi2005pof} investigated the streamwise development of $u_{rms}$ 
in a procession jet. 
As shown in their figure 18, 
the $u_{rms}$ of the precessing jet is asymmetric compared with a non-precession jet and exhibits two peaks, located tangentially `in front of' and `behind' the jet centre. 
The `front' peak is more pronounced than the `rear' peak. 
The similar behaviours between the jet under combined forcing 
and a precessing jet may suggest that the so-called combined forcing 
may have produced a precession jet. 
This suggestion is further corroborated by the downstream development 
of the centreline mean and fluctuating velocities 
$\overline U_{cl}/\overline U_j$ and $u_{cl,rms}/\overline U_j$ presented below. 

%--- Figure ------------------------------------------------------------
\begin{figure}
 	\centering
    \includegraphics[width=0.6\textwidth]{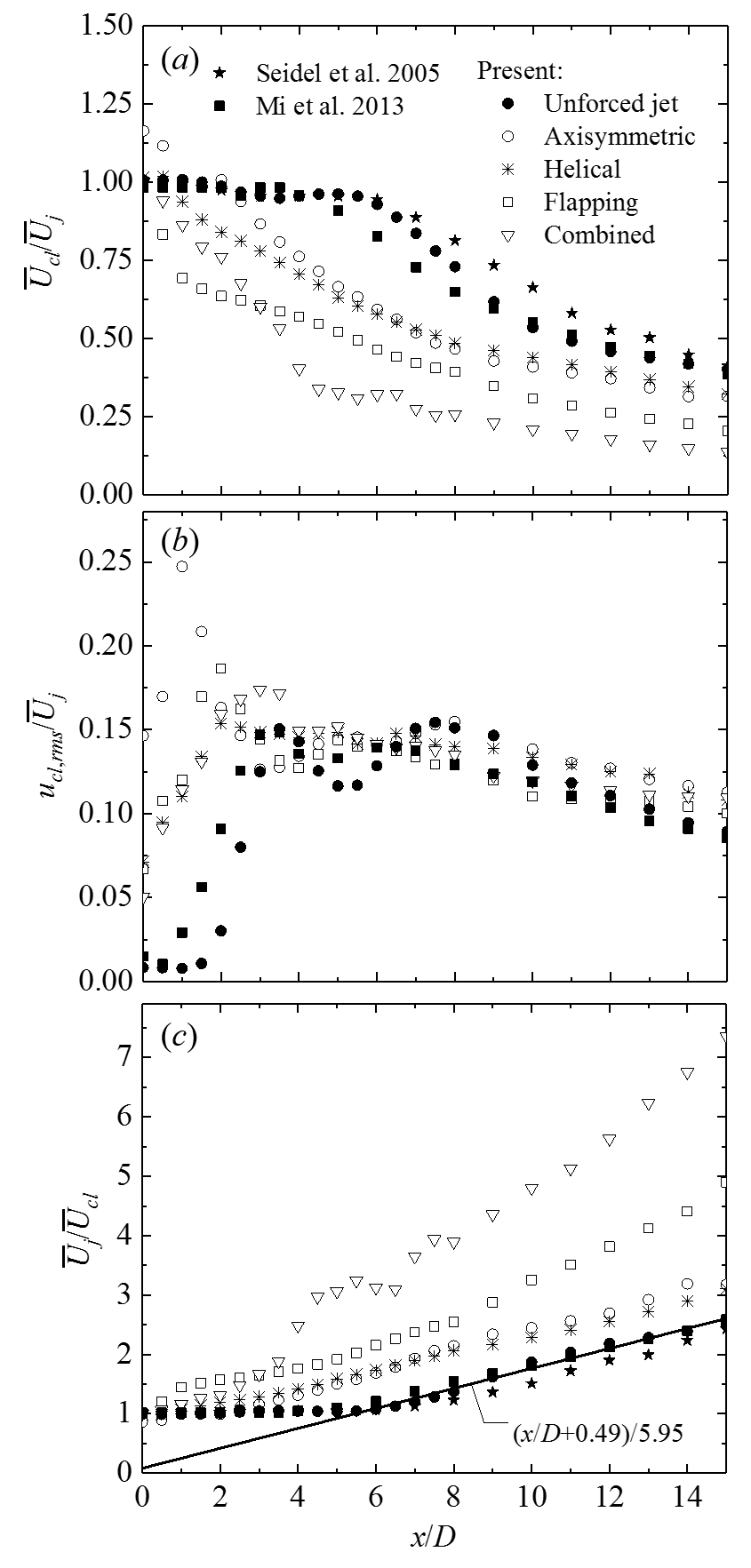}\\
    \caption{Streamwise variations of hot-wire measured centreline mean 
    and $rms$ velocities: 
    ($a$) $\overline U_{cl}/\overline U_j$, 
    ($b$) $u_{cl,rms}/\overline U_j$, 
    ($c$) $\overline U_j/\overline U_{cl}$.
    }
 	\label{Fig:centerline_profile}
 \end{figure}
%-----------------------------------------------------------------------

The variations in $\overline U_{cl}/\overline U_j$, $u_{cl,rms}/\overline U_j$ and $\overline U_j/\overline U_{cl}$ are presented in figure \ref{Fig:centerline_profile}
for various forcing modes as well as the unforced jet. 
It is worth pointing out that our contraction nozzle is extended 
by a 47 mm long smooth tube of the same diameter as the nozzle exit $D$ 
where the minijet assembly is mounted. 
In spite of this difference, the unforced jet displays the well-known features. 
Firstly, $f_0D/\overline U_j$ is 0.45, 
falling in the expected range 0.24 $\sim$ 0.64 
(e.g. \citet{GUTMARK1983pof}, \citet{Zhou2012AIAA}). 
Secondly, $\overline U_{cl}/\overline U_j$ and $\overline U_j/\overline U_{cl}$ 
(figure \ref{Fig:centerline_profile}$a, c$) agree both qualitatively 
and quantitatively with \citet{Mi2005pof} ($Re_D$ = 8050) 
and \citet{Seidel2005AIAA} ($Re_D$ = 8800) measurements. 
Following \citet{Todde2009eif}, 
we may fit the data to $\overline U_{cl}/\overline U_j$= $B[(x-x_0)/D]^{-1}$, 
where $x_0$ and $B$ denote the virtual origin and decay constant, respectively. 
Then, $x_0/D$ = - 0.49 and $\overline U_{cl}/\overline U_j$ decays 
at a rate of $x^{-1}$ beyond $x/D$ = 5, 
as reported by previous investigations (e.g. \citet{mi2001jfm}). 
Thirdly, the streamwise distribution of $u_{cl,rms}$ exhibits one peak 
at $x/D$ = 3.5 and another $x/D$ = 7.5 (figure \ref{Fig:centerline_profile}$b$). 
The former is connected to the breakdown of the primary ring vortices, 
and the latter is due to early transition to turbulence \citep{Mi2013pof}. 
Similar observations were made by \citet{Todde2009eif} (their figure 6) 
and by \citet{Mi2013pof} whose data are included in figure \ref{Fig:centerline_profile}($b$). 
Departures between the present and other's data are not unexpected 
in view of differences in, inter alia, experimental setup and $Re_D$ 
among the investigations. The potential core length of the unforced jet 
is approximately $5D$, beyond which $\overline U_{cl}$ appears dropping approximately linearly. 
Note that $\overline U_{cl}/\overline U_j$ for axisymmetric forcing 
exceeds 1.0 given $x/D$ $\le$ 2, higher than those of the other modes. 
This is due to an increase in the mass flow rate by 7.2\%. 
A similar observation is made by \citet{Seidel2005AIAA} who observed, 
with 16 minijets blowing, an increase in the centreline mean velocity 
near the jet exit. Under all control modes, 
$\overline U_{cl}/\overline U_j$  decays rapidly right from the beginning. 
The minijet actuation reduces $\overline U_{cl}/\overline U_j$ significantly 
at $x/D$ $\le$ 8 (figure \ref{Fig:centerline_profile}$a$), 
demonstrating the efficacy of minijet injections. 
This efficacy can be attributed to the earlier increase 
in the turbulence levels when the radial injections are present; 
a large turbulence level (figure \ref{Fig:centerline_profile}$b$) is correlated with a rapid drop in $\overline U_{cl}/\overline U_j$ (figure \ref{Fig:centerline_profile}$a$). 
Interestingly, the variations in $u_{cl,rms}/\overline U_j$ 
for helical and combined forcings are quite similar to each other, 
growing almost linearly first and then experiencing a small drop 
before fluctuating slightly around 0.15 (figure \ref{Fig:centerline_profile}$b$). 
This similarity is ascribed to the common feature of the two control modes, 
i.e. the swirling motion. 
However, it is combined forcing that maintains the rapid and 
linear growth further downstream, up to $x/D$ = 3 and retreats little 
by $x/D$ = 3.5. 
It is noted earlier in figure 14c that those peaks at $f/f_a$ = 1, 2, 3, 4 
and 5 in $E_u$ also remain pronounced up to $x/D$ = 3. 
This coincidence suggests that the excited coherent structures account 
for the linear growth in $u_{cl,rms}/\overline U_j$ 
(figure \ref{Fig:centerline_profile}$b$) and the rapid decay 
in $\overline U_{cl}/\overline U_j$ (figure \ref{Fig:centerline_profile}$a$).
In contrast, the rapid growth of $u_{cl,rms}/\overline U_j$ is only up to 
$x/D$ = 2 for helical forcing and even only up to 1 for axisymmetric 
and flapping forcings (figure \ref{Fig:centerline_profile}$b$). 
As such, $\overline U_{cl}/\overline U_j$ keeps decaying rapidly 
and almost linearly until $x/D$ = 4.5 for the combined forcing mode 
and remains well below other cases further downstream (figure \ref{Fig:centerline_profile}$a$). 
It is worth pointing out that the precessing jet is also characterized 
by a substantially faster decay than the non-precessing jet \citep{Mi2005pof}.

%--- Figure ------------------------------------------------------------
\begin{figure}
    \centering
    \includegraphics[width=0.5\textwidth]{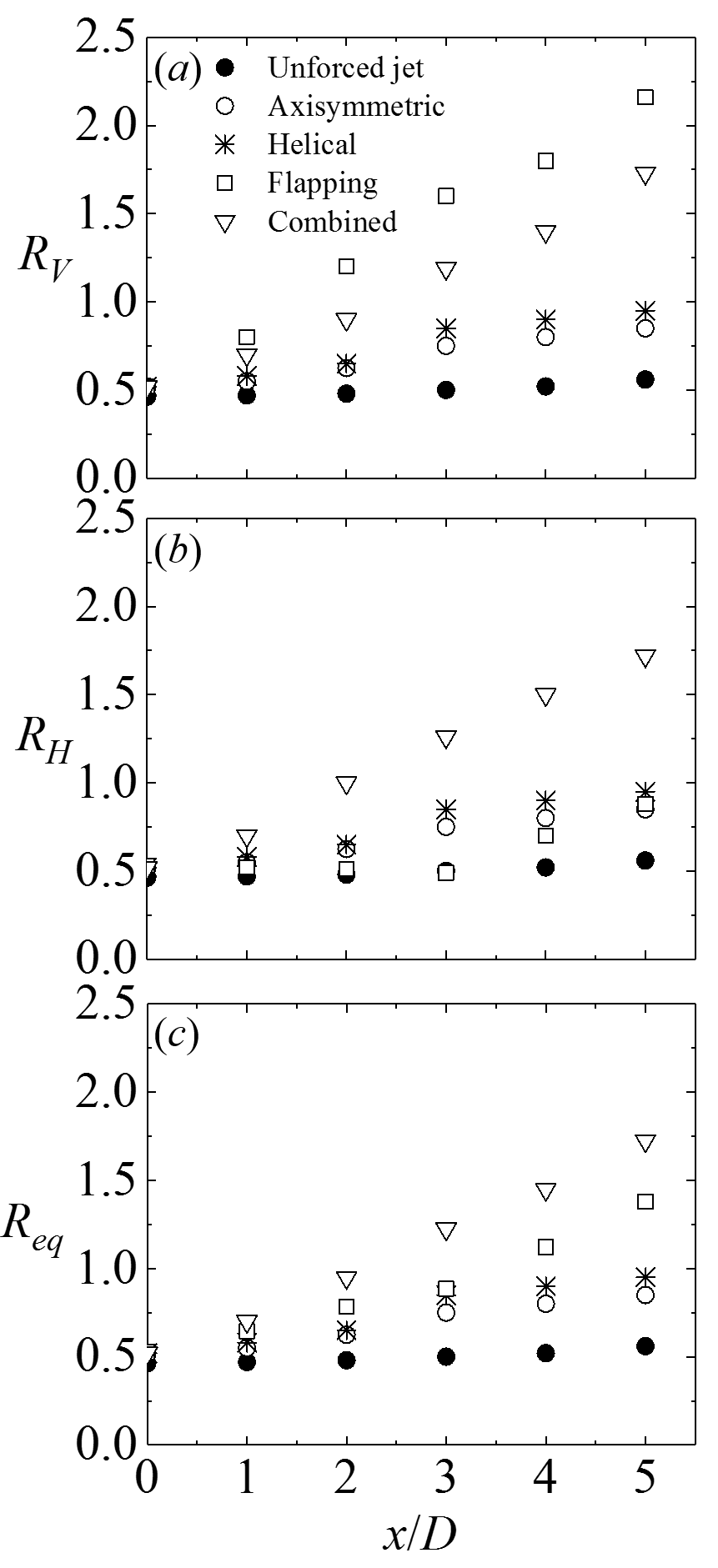}\\
    \caption{Mean-velocity half-widths at different $x/D$: 
    ($a$) $R_V$ in the $x-y$ plane, 
    ($b$) $R_H$ in the $x-z$ plane, and 
    ($c$) the equivalent half-radius $R_{eq}$
    }
    \label{Fig:half widths}
 \end{figure}
%-----------------------------------------------------------------------

Jet spreading rate and the overall entrainment rate may be well quantified 
by the downstream variation of the jet width (e.g. \citet{Zhou2012AIAA}). 
Following \citet{Hussain1989jfm}, we define an equivalent jet width 
by $R_{eq}$ = $[R_H R_V]^{0.5}$, where $R_H$ and $R_V$ 
denotes the mean-velocity half-widths in the $x-z$ and $x-y$ planes, respectively. 
The half-width is defined as the distance between the jet centreline 
and the location at which $\overline U = 0.5 \overline U_{cl}$. 
Figure \ref{Fig:half widths} shows the downstream evolution of $R_V$, $R_H$ and $R_{eq}$. 
While changing little for unforced jet, 
$R_V$, $R_H$ and $R_{eq}$ grow appreciably in controlled jet. 
Evidently, $R_{eq}$ is the largest for the combined mode, 
followed by flapping, helical and axisymmetric, 
though the latter two do not differ much. 
The results provide additional support for our choice of $J$ 
as a measure for the mixing efficacy.

%--- Figure ------------------------------------------------------------
\begin{figure}
 	\centering
    \includegraphics[width=1\textwidth]{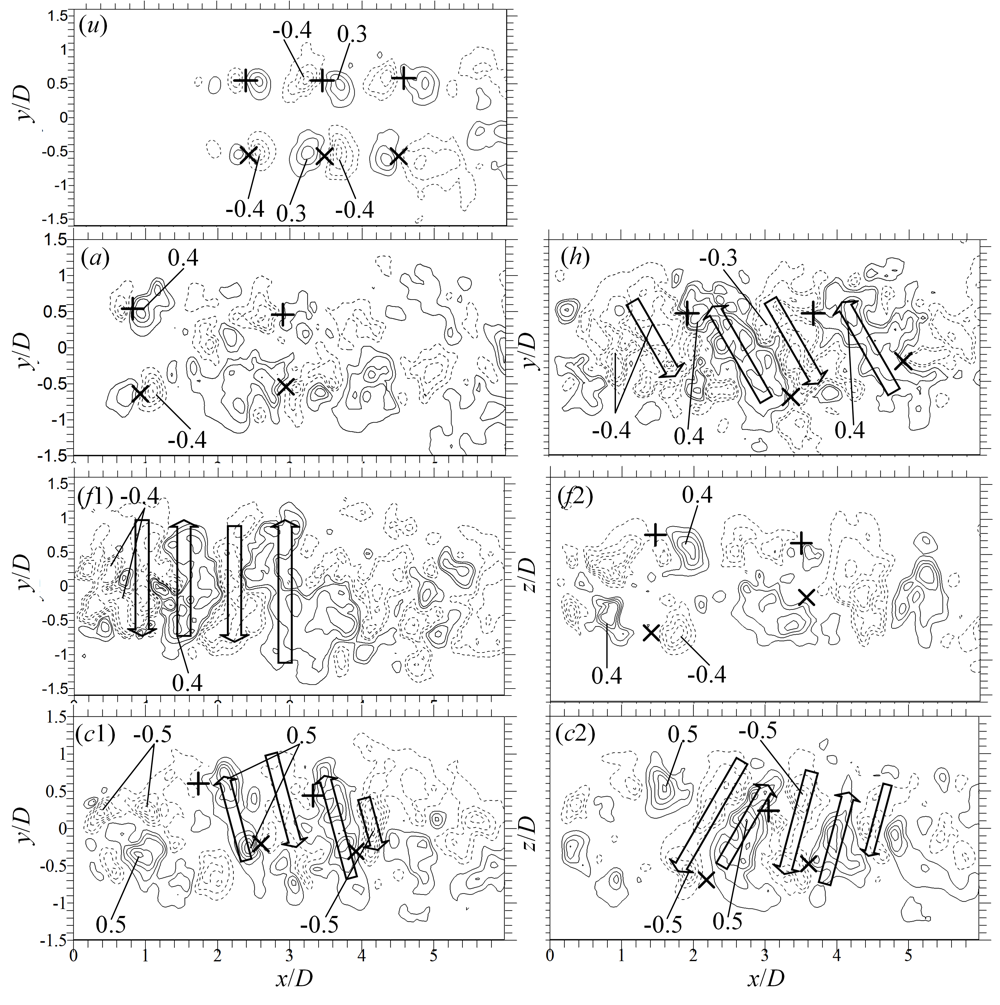}\\
    \caption{Iso-contours of typical instantaneous velocities 
    $V/\overline U_j$ or $W/\overline U_j$ in the $x-y$ and $x-z$ planes: 
    ($u$) unforced jet, 
    ($a$) axisymmetric, 
    ($h$) helical, 
    ($f$) flapping and 
    ($c$) combined forcing. 
    Contour interval = 0.1. 
    The lowest contour level is 0.1 for all plots. 
    Solid and broken contours represent the positive (upward motion) 
    and the negative (downward motion), respectively. 
    Symbols `+' and `$\times$' denote anti-clockwise and clockwise vortices, respectively. 
    The arrows indicate the moving direction of fluid slice.}
 	\label{Fig:PIV_streamwise}
 \end{figure}
%-----------------------------------------------------------------------

To understand further the predominant flow structures under the four forcings, 
we examine in figure \ref{Fig:PIV_streamwise} typical instantaneous $V$- or $W$-contours, 
measured using PIV, in the $x-y$ and $x-z$ planes. 
In the unforced jet, the positive and negative velocity concentrations occur 
in pair and are mirrored by another pair, though with swapped signs, 
on the other side of the centreline (figure \ref{Fig:PIV_streamwise}$u$). 
Apparently, the two pairs of velocity concentrations are associated 
with the two vortical structures, as indicated by symbols `+' and `$\times$', 
of one ring vortex. 
Axisymmetric forcing leads to the topologically unchanged velocity contours 
but a significantly increased size in the velocity concentrations 
which now start to occur in the near proximity of the nozzle exit (figure \ref{Fig:PIV_streamwise}$a$). 
Under helical forcing, the vortical structures above and below 
the centreline are stagger-arranged and their same-signed velocity contours 
are connected, forming inclined strips alternately in sign (figure \ref{Fig:PIV_streamwise}$h$). 
Under flapping forcing (figure \ref{Fig:PIV_streamwise}$f1$), 
the $V$-contours in the $x-y$ plane exhibit alternate upward and 
downward motions which are absent in the $W$-contours in the $x-z$ plane. 
Some interesting observations can be made for combined forcing. 
Firstly, the inclined strips of velocity contours, alternately in sign, 
are seen in both planes (figure \ref{Fig:PIV_streamwise}$c1, c2$). 
Secondly, the velocity concentrations reach in general the maximum 
contour level of $\pm$0.5, greater than those associated with other forcings 
($\pm$0.4), indicating a stronger spread/entrainment or mixing. 
Thirdly, the pattern of velocity contours and their maximum strength 
persists considerably further downstream, reaching $x/D$ $\approx$ 5,
whilst the upward and downward motions under flapping forcing disappear 
at $x/D$ $\approx$ 3, 
suggesting a prolonged entrainment and more thorough mixing. 
The observation is again internally consistent with the rapid drop 
until $x/D$ $\approx$ 4.5 in the centreline mean velocity decay (figure \ref{Fig:centerline_profile}$a$).

%--- Figure ------------------------------------------------------------
\begin{figure}
 	\centering
    \includegraphics[width=0.7\textwidth]{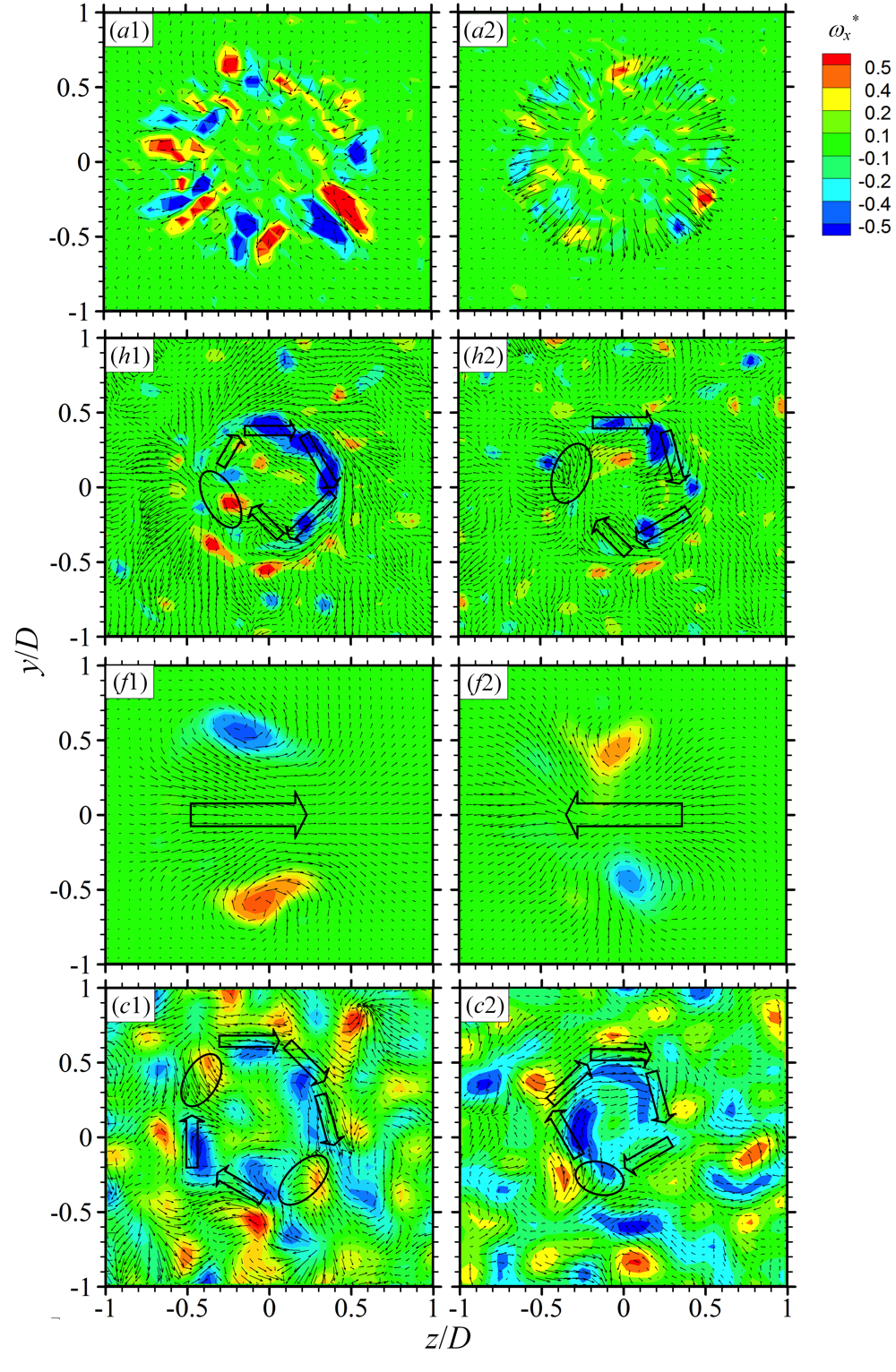}\\
    \caption{The typical PIV snapshots of instantaneous velocity vectors 
    in a cross-sectional plane of 
    ($u$)unforced jet,
    ($a$) axisymmetric, 
    ($h$) helical, 
    ($f$) flapping and 
    ($c$) combined forcings at $x/D = 0.25$. 
    The open arrow represents the motion of jet fluid. 
    The elliptic contour indicates where fluid is entrained into jet core area.
    }
 	\label{Fig:PIV_cross}
 \end{figure}
%-----------------------------------------------------------------------

More insight may be gained into the predominant flow structures under control 
by examining typical instantaneous velocity vectors superimposed 
with the corresponding iso-contours of streamwise vorticity 
$\omega_x^\star$ = $\omega_x D/\overline U_j$ 
in the $y-z$ plane at $x/D$ = 0.25 (figure \ref{Fig:PIV_cross}). 
Under axisymmetric forcing (figure \ref{Fig:PIV_cross}$a1, a2$), 
the vectors show the inward or outward motions associated 
with the ring-like structures, which are axisymmetric and highly repeatable. 
There are six pairs of counter-rotating $\omega_x^\star$ concentrations arranged about the centre (figure \ref{Fig:PIV_cross}$a$), 
apparently generated by the six axisymmetrically placed minijets. 
The rotational motion under helical forcing is evident and the fluid 
moves inward along the circumference (figure \ref{Fig:PIV_cross}$h1, h2$). 
The core region appears rather stagnant. 
The phase of injecting minijets is clockwise incremented by $60^\circ$ 
(figure \ref{Fig:Actuation}$h$), 
producing a corkscrew type of structure (figure \ref{Fig:Crossplane}$h1-h6$), 
as shown by \citet{Koch1989pof}. 
Figure \ref{Fig:PIV_cross}($f1, f2$) shows the cross-flow motion 
that switches from one direction at one moment to the opposite 
at another under flapping forcing, accompanied 
by one pair of counter-rotating $\omega_x^\star$ concentrations, 
as shown by \citet{Yang2016jfm}. 
The velocity vectors in Figure \ref{Fig:PIV_cross}($c1, c2$) 
exhibit the clockwise rotational motion under combined forcing. 
The area of rotational motion, as indicated by the arrows 
in figure \ref{Fig:PIV_cross}$c1$ is in general larger than 
that under helical forcing (figure \ref{Fig:PIV_cross}$h1-h2$). 
Note that ambient fluid may be entrained into the jet core area 
from various circumferential locations, e.g. the upper left and 
lower right corners as indicated by the elliptic contours 
in figure \ref{Fig:PIV_cross}$c1$, 
while under helical forcing ambient fluid comes into the core area largely 
from only one location as highlighted by the elliptic contours 
(figure \ref{Fig:PIV_cross}$h1, h2$). 
Furthermore, there are many vorticity concentrations of both signs 
in figure \ref{Fig:PIV_cross}($c1-c2$). 
The core area is dominated by the vorticity concentrations of negative sign, 
while the region surrounding the core is populated with those of both signs. 
This is very different from flapping forcing where there is only 
one pair of opposite-signed vorticity concentrations. 
This is also markedly different from helical forcing (figure \ref{Fig:PIV_cross}$h1, h2$) where the cross-sectional plane is characterized 
by the vorticity concentrations of a single sign. 
The observations reconfirm that the combined forcing mode 
is associated with a much better mixing and furthermore probably 
also small-scale mixing.

\subsection{Insight into the 3D flow structure}
\label{ToC:Insight into the 3D flow structure}

%--- Figure ------------------------------------------------------------
\begin{figure}
 	\centering
    \includegraphics[width=0.95\textwidth]{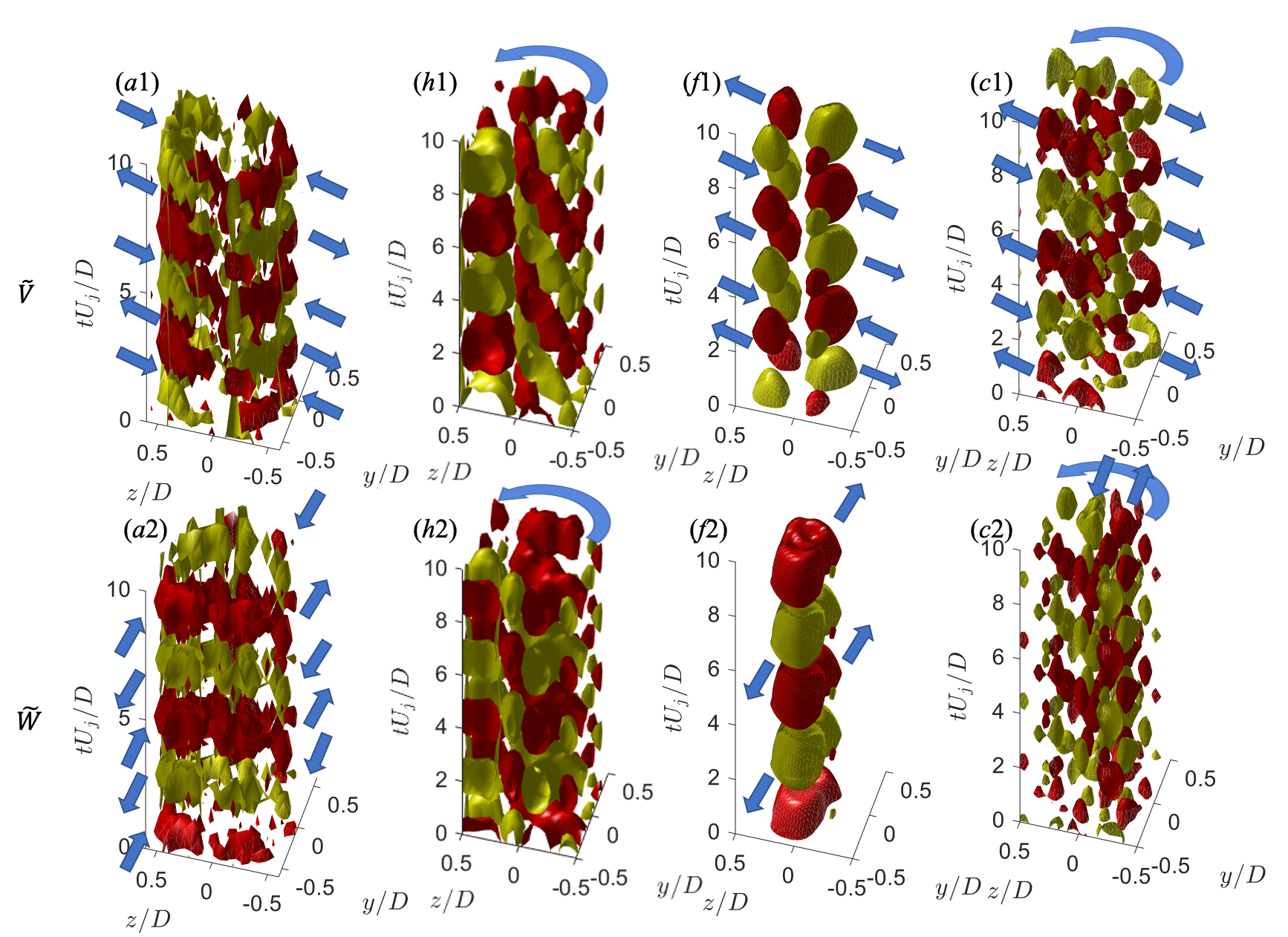}\\
    \caption{Identification of large coherent structures 
    by iso-surfaces of phase-averaged velocity for 
    ($a$) axisymmetric 
    ($h$) helical 
    ($f$) flapping 
    ($c$) combined forcings at $x/D = 0.25$. 
    The blue arrow indicates the direction of jet flow.
    }
 	\label{Fig:Cartoon}
 \end{figure}
%-----------------------------------------------------------------------

To gain insight into the three-dimensional (3D) flow structure, 
we deploy a linear stochastic estimation method to 
reconstruct the predominant flow structure from the PIV data, 
captured in the $y-z$ plane at $x/D$ = 0.25 
with a total of 200 images. 
This technique is introduced in detail by e.g. 
\citet{adrian1988jfm} and is briefly described below.

Let $q(y, z, t)$ be a velocity component in the cross plane of $x/D$ = 0.25 
recorded at constant sampling rate with time step $\triangle t$, i.e. $t_m$ = $m \triangle t$, $m$ = $1 ,2\ldots, M$, 
where $M$ denotes the total number of the PIV snapshots. 
Use $q^m(y, z)$ = $q(y, z, t^m)$ to denote the corresponding snapshots. 
The mean flow is given by
 
\begin{equation}
\overline q(y,z) = \frac{1}{M}\sum_{m=1}^M q^m(y,z).
\label{Eqn:velocity field}
\end{equation}
The oscillatory actuation response is the corresponding Fourier component 
of the fluctuation $q' = q - \overline q$. 
The cosine and sine contributions read

\begin{subequations}
\begin{eqnarray}
&q_1(y,z)= \frac{2}{M} \sum_{m=1}^{M} \cos\phi^m[q^m(y,z)-\overline q(y,z)]
\label{Eqn:cos component}   \\
&q_1(y,z)= \frac{2}{M} \sum_{m=1}^{M} \sin\phi^m[q^m(y,z)-\overline q(y,z)]
\label{Eqn:sin component}        \\
&\phi^m=2\pi f_a t^m.
\label{Eqn:phi}
\end{eqnarray}
\label{Eqn:velocity component}
\end{subequations}%
Thus, the periodic flow response may be given by

\begin{equation}
\widetilde q(y,z, t) = \overline q(y,z) +  q_1(y,z)cos\phi(t) + q_2(y,z)sin\phi(t), \phi=2\pi f_a t.
\label{Eqn:reconstruction}
\end{equation}
The residual of this phase-averaged flow $\widetilde q$ consists of 
higher harmonics and an uncorrelated stochastic contribution. 
Note the actuation commands $b_i(t)$ are the functions of the pointer 
($cos \phi$, $sin \phi$). 
Hence, the temporal Fourier component optimally represents 
the actuation response and no flow-intrinsic phase needs to be constructed.

The iso-surfaces of reconstructed $\widetilde V$ and $\widetilde W$ 
are presented in figure \ref{Fig:Cartoon} for four control modes. 
The flow structures of axisymmetric, helical and flapping forcing modes 
show excellent agreement with the control laws (figure \ref{Fig:Actuation}). 
Under axisymmetric forcing (figure \ref{Fig:Cartoon}$a1-a2$), 
$\widetilde V$ and $\widetilde W$ indicate clearly that jet fluids 
on the two sides of the centreline move 
either inward or outward simultaneously (figure \ref{Fig:Cartoon}$a$). 
Under helical forcing (figure \ref{Fig:Actuation}$h1-h6$), 
the helical motion is evident (figure \ref{Fig:Cartoon}$h$). 
For flapping forcing, as indicated by the iso-surfaces of $\widetilde W$, 
one sector of fluid moves in one direction for one instant and 
the adjacent sectors move in the opposite direction (figure \ref{Fig:Cartoon}$f2$). 
On the other hand, the iso-surfaces of $\widetilde V$ is anti-symmetrical 
about the centreline (figure \ref{Fig:Cartoon}$f1$). 
The observations are fully consistent with the characteristics of 
the flapping motion documented in e.g. \citet{Yang2016jfm}. 
The results provide a validation for the presently reconstructed $\widetilde V$ and $\widetilde W$. 
For the combined mode (figure \ref{Fig:Cartoon}$c$), 
the flow structure appears much more complicated. 
However, the swirling motion is still discernible. 
Furthermore, the iso-surfaces of $\widetilde V$ and $\widetilde W$ 
show unequivocally the occurrence of many more small parcels of fluids, 
suggesting a much better jet mixing than all other flow modes, 
which is fully consistent with the finding from figure \ref{Fig:LearningCurve} as well as figure \ref{Fig:PIV_cross}($c$).

\section{Conclusions and outlook}
\label{ToC:Conclusions}

An artificial intelligence (AI) control system has been developed 
for the control of turbulence. 
The system consists of a control plant, a sensing unit, 
an actuation unit, and a control or `thinking' unit. 
A round jet (control plant) is manipulated to 
illustrate the potential of this system. 
Two hot-wires are deployed for sensing online 
the information on the turbulent jet. 
The control unit deploys a linear genetic programing, 
and six independent unsteady radial minijets 
placed around the nozzle prior to the issue of main jet 
for executing control laws generated from the control unit. 
The search space for control laws is extremely large, 
including the minijet number $N$, geometric configuration, 
frequency $f_a$, duty cycle $\alpha$ and phase shift $\varphi_{ij}$ 
($i$, $j$ = $1, 2, \dots, 6$) between minijets. 
The target is to maximize the decay rate $K$ of 
the jet centreline mean velocity, 
which is correlated with the mixing rate.

It has been demonstrated that the AI control system 
can learn automatically how to optimize 
the spatially distributed actuators 
and thus the turbulent jet for the targeted cost. 
Like virtually all other control strategies of nonlinear dynamics, 
AI control solutions do not come with a proof of global optimality. 
Yet, the results for jet mixing optimization demonstrate 
several highly desirable features. 
First, AI control has identified a few typical and well-known control laws, 
i.e. axisymmetric, helical and flapping forcings, 
in its learning process in the order of increased control performances, 
and eventually converged to an unexpected spatio-temporal forcing, 
referred to as the combined mode, which has never been reported previously. 
The learning time of 1100 individuals or 2 hours wind-tunnel testing is remarkably short 
for such a complex solution. 
It is worth pointing out that the presently developed AI control method 
may not suit for numerical simulation where faster learners, 
e.g. the reduced-order-model-based models, would be required 
(once the winning control mechanism is known) to reduce significantly 
the testing time for N-S-based simulations. 
Inspirations may be gained from control studies of a turbulent boundary layer \citep{sasaki2020jfm} 
or turbulent separation \citep{nair2019jfm}. 
Second, the AI-learned combined mode is reproducible with other initial generations. 
The control laws may analytically differ but produce almost identical actuation commands. 
Third, the parameters of the underlying linear genetic programming are taken verbatim from \citet{Duriez2016mlc} and 
were already proven useful in many other experiments. 
No sensitive dependence on the parameters has been observed so far 
and AI control can be expected to yield near-optimal results 
in its first application to a new plant. 
Finally, the unique advantage of the AI control over conventional techniques, 
be open or closed-loop and linear or nonlinear, 
is its capability to find the apparently global optimum solution 
when the search space for control laws is extremely large. 
This advantage will be lost in case of a single actuator 
involving few control parameters such as frequency and duty cycle 
(e.g. \citet{Fan2017AIAA}).

The control landscape is studied by examining the feature vectors of generations $n$ = 1, 2, 5 and 11. 
Several interesting findings are made. 
Firstly, jet mixing benefits from the increasing asymmetry of forcing in the learning process. 
Secondly, the feature vectors appear randomly distributed 
at $n$ = 1 but evolve gradually to three distinct regions with increasing $n$ (figure 12), 
characterized by axisymmetric, helical and flapping forcings, respectively. 
While the region of axisymmetric forcing contracts to one single point, 
the other two shrink essentially to two curves in the evolution process. 
Intriguingly, the converged globally optimized control law 
at $n$ = 11 takes place in the region of helical forcing 
which is now populated with either helical or combined mode. 
Thirdly, the feature coordinate $\gamma_1$ is found to be correlated with the degree of asymmetric forcing, 
while $\gamma_2$ is linked to the number of simultaneously injecting minijets. 
The best control law is characterized by the largest $\gamma_2$
for axisymmetric forcing and largest $\gamma_1$ 
for the other two forcings. 
However, the globally optimized control law (combined mode) 
is associated with neither the largest $\gamma_1$ nor largest $\gamma_2$.

This combined forcing has produced a novel turbulent flow structure 
characterized by strong oscillation and swirling motions, 
along with the generation of mushroom-like structures, 
all acting to enhance jet mixing. 
As a result, the combined mode vastly outperforms 
the well-known optimal benchmark forcings, increasing the entrainment or mixing rate by $54\%$, $47\%$ and $28\%$ compared with 
the axisymmetric, helical and flapping forcings, respectively. 
Extensive measurements are conducted in three orthogonal planes to understand this novel flow structure, 
which is compared with the flow structures under the benchmark axisymmetric, helical and flapping forcings. 
It has been found that this flow structure is characterized by a number of features, including 
(1) the helical motion, 
(2) 3D oscillating jet column, 
(3) strong coherent structures, 
(4) asymmetrical cross-flow distributions of mean and fluctuating velocities, 
the latter displaying twin-peak behavior in every plane through the $x$-axis, 
(5) spiral behavior of the swirling jet centre, 
(6) changing speed of swirling, 
(7) spectral phase shift by $\pi$ over a considerable frequency band about $f/f_a$ = 1 
between two simultaneously captured hotwire signals placed 
at ($x/D$, $y/D$) = (1, $\pm$ 0.3), and 
(8) many vorticity concentrations of opposite signs over the entire cross-sectional jet plane, 
which is distinct from the flow structures under other forcings and suggests better and smaller scale mixing. 
It is further found that features (1) through (5) resemble those of a precession jet generated by a passive device.

We expect that AI control will be commonly applied to 
discover the unknown winning nonlinear actuation mechanism 
of multi-input multi-output flow control experiments in the very foreseeable future. 
Conventional model-based or model-free control design may be 
then deployed to refine and optimize the AI-based actuation mechanism, 
provided the control law is sufficiently simple. 
AI control may be improved in numerous aspects. 
Examples include a human-interpretable control law, 
an increased learning speed in experiment, 
robustness against varying operating conditions 
and the inclusion of prior knowledge and expectations of control laws. 
One can safely assume that AI will be an essential tool 
in future turbulence control applications, 
just as AI is indispensable in robotics now.

\begin{acknowledgements}
YZ wishes to acknowledge support given to him from NSFC through 
grants 11632006, 91752109 and 91952204. 
This work is supported by the French National Research Agency (ANR) 
via the grants ANR-11-IDEX-0003-02 (iCODE), `ACTIV ROAD' 
and `FlowCon', and by the OpenLab Fluidics consortium 
(Fluidics@poitiers) of PSA Peugeot-Citro\"e n and Institute Pprime.
\end{acknowledgements}

\begin{appendix}
\section{Control laws}
\label{ToC:Appendix}

AI control discovers four typical actuations, as given below. 
The best individual of the first generation or stage 1 
is characterized by an axisymmetric control law:

\begin{equation}
b_1 = b_2 = b_3 = b_4 = b_5 = b_6 = -0.832 + \sin (\omega_at + 4/6\pi).
\label{Eqn:generation_1}
\end{equation}

Stage 2 starts with the second generation when the 
AI control discovers a helical forcing:

\begin{subequations}

\begin{eqnarray}
&& b_1 = \sin (\omega_at + 4/6\pi) - 0.145, 
\label{Eqn:generation_2_b1}   \\
&& b_2 = -0.347 \sin \omega_at,
\label{Eqn:generation_2_b2}        \\
&& b_3 = [\sin (\omega_at + 8/6\pi) + \sin (\omega_at + 8/6\pi)^2 
+ \sin (\omega_at + 2/6\pi)^2] \nonumber\\
&&  \sin (\omega_at + 8/6\pi) ,
\label{Eqn:generation_2_b3} \\
&& b_4 = 2\sin (\omega_at + 10/6\pi) [(\sin (\omega_at)^2 - 
\sin (\omega_at + 2/6\pi) \nonumber\\ 
&& (\sin (\omega_at)^2 - \sin (\omega_at + 2/6\pi))] , 
\label{Eqn:generation_2_b4} \\
&& b_5 = 1/ (-0.347 + \sin \omega_at) + \sin \omega_at,
\label{Eqn:generation_2_b5} \\
&& b_6 = -0.354 \sin (\omega_at + 8/6\pi). 
\label{Eqn:generation_2_b6}
\end{eqnarray}

\label{Eqn:generation_2}
\end{subequations}%

Flapping forcing takes place in stage 3, starting from the fifth generation:

\begin{subequations}

\begin{eqnarray}
&& b_1 = b_2=b_3= - 0.811 + \sin (\omega_at + 2/6\pi), 
\label{Eqn:generation_5_b1}   \\
&& b_4 = b_5=b_6=  - 0.782 - \sin (\omega_at + 2/6\pi).
\label{Eqn:generation_5_b4}     
\end{eqnarray}

\label{Eqn:generation_5}
\end{subequations}%

The learning process converges to a complex control law 
in the eleventh generation:

\begin{subequations}

\begin{eqnarray}
&& b_1 = ((\sin (\omega_at + 6/6\pi)/\sin (\omega_at + 4/6\pi))2/\sin \omega_at - \sin \omega_at)/ (\sin \omega_at)^4, 
\label{Eqn:generation_11_b1}   \\
&& b_2=((((\sin (\omega_at + 10/6\pi)/(\sin \omega_at)^2 - 1/\sin \omega_at - \sin \omega_at)/\nonumber\\
&& (\sin \omega_at)^2 + \sin \omega_at)/(\sin \omega_at)^3/\sin (\omega_at + 10/6\pi) - \sin \omega_at)/(\sin \omega_at)^2/ \nonumber\\ 
&& \sin (\omega_at + 8/6\pi)/ -0.811/( \sin (\omega_at + 10/6\pi)/(\sin \omega_at)^2 - \sin \omega_at)/(\sin \omega_at)^2/\nonumber\\
&& \sin (\omega_at + 8/6\pi)/ -0.811 ,
\label{Eqn:generation_11_b2} \\
&& b_3 = -811/(-0.811+ \sin (\omega_at + 2/6\pi) + (0.482 - \sin (\omega_at + 10/6\pi))^2/\nonumber\\
&& (0.482 - \sin (\omega_at + 10/6\pi)),
\label{Eqn:generation_11_b3} \\
&& b_4 = \sin (\omega_at+10/6\pi) - 2\sin (\omega_at+ 2/6\pi)-0.223+(\sin (\omega_at + 2/6\pi)- \nonumber\\
&& \sin (\omega_at + 10/6\pi))^2,
\label{Eqn:generation_11_b4} \\
&& b_5 = ((\sin (\omega_at + 10/6\pi)/(\sin \omega_at)^2 - \sin \omega_at)/(\sin \omega_at)^2 + \sin \omega_at)/\nonumber\\
&& (-0.782 + \sin (\omega_at + 2/6\pi)  - \sin (\omega_at + 10/6\pi)/(\sin \omega_at)^2/\sin (\omega_at + 8/6\pi) + \nonumber\\
&& (\sin (\omega_at + 10/6\pi)/(\sin \omega_at)^2 - \sin \omega_at)/(\sin \omega_at)^2/(-0.782 + \sin (\omega_at + 2/6\pi) + \nonumber\\
&& (-0.782+\sin (\omega_at + 2/6\pi))^2)/((\sin (\omega_at + 10/6\pi)/(\sin \omega_at)^2 - \sin \omega_at)/(\sin \omega_at)^2+ \nonumber\\
&& \sin \omega_at)/(-0.782 + \sin (\omega_at + 2/6\pi) -\sin (\omega_at + 10/6\pi)/(\sin \omega_at)^2)/\nonumber\\
&& (\sin (\omega_at + 8/6\pi)+ \sin (\omega_at + 10/6\pi)/(\sin \omega_at)^2 - \sin \omega_at)/(\sin \omega_at)^2/\nonumber\\
&& (-0.782+ \sin (\omega_at + 2/6\pi) - \sin (\omega_at + 10/6\pi))/(\sin (\omega_at + 8/6\pi) +\nonumber\\
&& (\sin (\omega_at + 10/6\pi)/(\sin \omega_at)^2 - \sin \omega_at)/(\sin \omega_at)^2) +(((\sin(\omega_at + 10/6\pi)/ \nonumber\\
&&(\sin \omega_at)^2-\sin \omega_at)/ (\sin \omega_at)^2 + \sin \omega_at)/((-0.782 + \sin (\omega_at + 2/6\pi) - \nonumber\\
&& \sin (\omega_at + 10/6\pi)/(\sin \omega_at)^2)/\sin (\omega_at + 8/6\pi) + (\sin (\omega_at + 10/6\pi)/ \nonumber\\
&& (\sin \omega_at)^2 - \sin \omega_at)/(\sin \omega_at)^2)), 
\label{Eqn:generation_11_b5} \\
&& b_6 = 2 \sin (\omega_at + 10/6\pi) - \sin (\omega_at +6/6\pi). 
\label{Eqn:generation_11_b6}
\end{eqnarray}

\label{Eqn:generation_11}
\end{subequations}%

\end{appendix}

%\enlargethispage{10mm} % EXCEPTION

\bibliographystyle{jfm}
% Note the spaces between the initials
\bibliography{Main_Fan}

\end{document}